\documentclass{article}

\usepackage{arxiv}

\usepackage[utf8]{inputenc} 
\usepackage[T1]{fontenc}    
\usepackage{url}            
\usepackage{booktabs}       
\usepackage{amsfonts}       
\usepackage{nicefrac}       
\usepackage{microtype}      
\usepackage{graphicx}
\usepackage{doi}
\usepackage[all]{xy}
\usepackage{setspace}
\usepackage{xcolor}
\usepackage{tikz}
\usetikzlibrary{cd}
\usepackage{soul}
\usepackage{comment}

\title{A geometric one-fluid model of superfluid helium-4}

\author{Nadine Suzan Cetin\\
 Mathematical Institute, Faculty of Mathematics and Physics, Charles University,\\ 
 Sokolovsk\'{a} 83, 18675 Prague, Czech Republic
\And Michal Pavelka \\
 Mathematical Institute, Faculty of Mathematics and Physics, Charles University,\\ 
 Sokolovsk\'{a} 83, 18675 Prague, Czech Republic\\
Corresponding author: pavelka@karlin.mff.cuni.cz\\ 
\And Emil Varga\\
Faculty of Mathematics and Physics, Charles University, Ke Karlovu 3, 121 16
Prague, Czech Republic
}

\date{}

\newcommand{\shorttitle}{A geometric model for superfluid helium-4}

\hypersetup{
pdftitle={Geometric one-fluid model of superfluid helium-4},
pdfsubject={}
pdfauthor={N. Cetin, M. Pavelka, E. Varga}
pdfkeywords={superfluids, Hamiltonian mechanics}
}

\usepackage{amsmath,amssymb,amsfonts,latexsym,float,graphics,epsfig}
\newcommand{\rr}{\mathbf{r}}
\newcommand{\qq}{\mathbf{q}}
\newcommand{\pp}{\mathbf{p}}

\newcommand{\xx}{\mathbf{x}}
\newcommand{\vv}{\mathbf{v}}

\newcommand{\mm}{\mathbf{m}}

\newcommand{\ee}{\mathbf{e}}
\newcommand{\diff}{\mathrm{d}}
\newcommand{\dd}{\diff}

\newcommand{\oomega}{\boldsymbol{\omega}}
\newcommand{\FF}{\mathbf{F}}
\newcommand{\RRR}{\mathfrak{R}}
\newcommand{\TT}{\mathbf{T}}
\newcommand{\cc}{\mathbf{c}}
\newcommand{\cdv}[1]{\frac{\mathrm{D} #1}{\mathrm{D} t}}
\newcommand{\DD}{\mathbf{D}}
\newcommand{\vns}{\vv_{ns}}
\newcommand{\LL}{\mathbf{L}}
\newcommand{\LLL}{\mathcal{L}}
\newcommand{\rrr}{\mathfrak{r}}

\newcommand{\drr}{\dd\rr\,}
\newcommand{\dpp}{\dd\pp\,}

\begin{document}
\maketitle

\begin{abstract}
A standard description of superfluid helium-4 is based on the concept of two components (superfluid and normal), which leads to the so called two-fluid models. However, as there are no two kinds of atoms in helium-4, the two components can not be separated. Superfluid helium-4 is not a mixture of two components, being rather a single fluid with two motions. Here, we present a geometric one-fluid model of superfluid helium-4, which is based on the Hamiltonian formulation of fluid mechanics. The model is derived from the kinetic theory of excitations (treated as an ideal Bose gas under the temperature $1.3K$) and average particle motions. It can be simplified to the Hall-Vinen-Bekharevich-Khalatnikov (HVBK) two-fluid model, where it removes one fitting parameter from the HVBK model, but it also gives extra terms beyond the HVBK model. Actually, we show that the two-fluid models are problematic in case of higher counter-flow velocities, where the usual splitting of total momentum to the superfluid and normal component becomes impossible. Finally, we show how vortex line density may be added to the state variables. The one-fluid model can be seen as a generalization of the two-fluid models that is geometrically consistent, fully compressible, with non-zero superfluid vorticity, and compatible with classical experiments.
\end{abstract}

\section{Introduction}
Although superfluid helium-4 is usually described as a mixture of a superfluid component and a normal component (the so called two-fluid models), it is not a mixture of two fluids. It does not consist of two kinds of atoms. Instead, it should be perceived as one fluid that exhibits two motions \cite{landau6}.
Here, we present a Hamiltonian one--fluid model of superfluid helium-4 with foundations in quantum theory, which generalizes the two--fluid models and shows their limitations.

Superfluid helium-4 (or Helium II) is a quantum liquid exhibiting both classical hydrodynamic and quantum behavior. A widely recognized model for the description of the flow of Helium II was introduced by Tisza and Landau in the first half of the 20th century \cite{tisza,Landau1941}. This model has become known as the two--fluid model, and today there are many extensions of the basic idea that Helium II consists of two components. One of the components is called normal with density $\rho_n$ and velocity $\vv_n$ and is governed by a Navier--Stokes-type set of equations. The other component is called superfluid, it has density $\rho_s$ and velocity $\vv^s$, and it is governed by an Euler-type set of equations \cite{ketterle}. The latter is connected with the proportionality between $\vv^s$ and the gradient of the macroscopic wave function within the Gross--Pitaevskii theory of the weakly interacting Bose gas \cite{gross,pitaevski,landau9,davis2001}. The Schrödinger equation can be then used for instance for the description of vortex rings in superfluid helium-4 \cite{berloff1999,berloff2007} or vortex nucleation \cite{muller-krstulovic}.

The Landau-Tisza two-fluid model was then extended by adding interactions between the two components, the so--called mutual friction force. This extension, where also the so-called quantized vortices and their average dynamical behavior can be described, is called the Hall-Vinen-Bekarevich-Khalatnikov (HVBK) model \cite{HallVinen56,BK1961}. For a thorough overview and critical discussion see, for example, \cite{nemirovskii2013, donnelly1999}.

Although the two--fluid models describe the physical behavior of Helium II well in many situations, they suffer from some principial insufficiencies. One problem is that the model takes parts of other theories (hydrodynamics, thermodynamics, many-body condensed matter theories), and puts them together in such a way that there is good agreement with experimental data, instead of deriving all from the first principles \cite{wong2008}.

Another pitfall is that it seems impossible to separate the normal component from the superfluid component and for instance weigh them, as the liquid has only one kind of atoms, just as Landau \& Lifshitz emphasize \cite{landau6}.

This results in the problem that for higher counter-flow velocities, the total momentum can no longer be split into the sum $\rho_s\vv^s+\rho_n\vv_n$, where we have the densities and velocities of the two components. The two--fluid model thus breaks down in this regime, see Section \ref{sec.beyond}.

The geometric one--fluid model presented in this publication is a Hamiltonian model of Helium II with only one mass density as a state variable, on top of the total momentum, entropy, and superfluid velocity. The basic idea is to describe the overall liquid by the dynamics of the matter part (fluid cells or average atom motions) and the dynamics of excitations (phonons and rotons) within one model. Both parts, matter and excitations, then behave hydrodynamically. Although the model is equipped with two velocities, similarly as the two-fluid models, we call it a one--fluid model, since it does not use the concept of two components in superfluid helium-4.

Note that we do not address helium-3, which also exhibits superfluidity, is composed of fermions, and has many peculiar properties \cite{volovik-droplet}, but we leave helium-3 for a future work. 

The plan of this manuscript: Section \ref{sec:model comparison} recalls various approaches towards modelling of Helium II. Section \ref{sec.one} contains the derivation and discussion of the geometric one-fluid model. Finally, Section \ref{sec.VLD} indicates how the vortex line density may be introduced as a state variable. 

\section{Existing models of and theoretical approaches to superfluid Helium 4} \label{sec:model comparison}

The landscape of models describing superfluid helium-4 is vast and hard to fully survey. Here, we present only some of the theories that are most related to the geometric one-fluid model. 


\subsection{The Gross--Pitaevskii (GP) theory}

The GP theory is a very general quantum theory of condensed matter physics for many interacting particles \cite{gross,pitaevski,brachet}. The starting point is a many--body Hamiltonian of condensed matter physics in the second quantization,
\begin{equation} \label{GP Hamiltonian}
    \hat H= \int \drr \hat{\psi}^\dagger (\mathbf r) \frac{\hat p^2}{2m}\hat{\psi} (\mathbf r) + \frac{1}{2}\int \drr  \int \drr'  \hat{\psi}^\dagger (\mathbf r') \hat{\psi}^\dagger (\mathbf r) \hat U(\mathbf r-\mathbf r')  \hat{\psi} (\mathbf r')\hat{\psi} (\mathbf r)
\end{equation}
where $\hat{\psi} (\mathbf r)$ and $\hat{\psi}^\dagger (\mathbf r)$ are field operators describing the annihilation and creation, respectively.

Under certain assumptions (Bogoliubov approximation \cite{landau9}), the macroscopic wave function $\phi(\mathbf r)$ has the following form 
\begin{equation} \label{macroscopic WV}
    \phi(\mathbf r,t)=\sqrt{n(\mathbf r,t)}\exp{(i\varphi(\mathbf r,t))},
\end{equation}
which describes all condensed particles by the particle density $n(\mathbf r,t)$. The coherence comes from the phase $\varphi(\mathbf r,t)$, which yields the same phase velocity for all condensed particles. The GP equation is then 
\begin{equation} \label{GP equation}
i\hbar \partial_t \varphi(\mathbf r,t)=\left( -\frac{\hbar^2}{2m}\Delta +g |\phi\mathbf (\rr,t)|^2 +U(\rr)\right) \varphi(\mathbf \rr,t).
\end{equation}
Finally, separating the spatial and temporal dependencies and decomposing Equation \eqref{GP equation} into the real and imaginary parts, a continuity equation and an Euler--like equation emerge (similarly as in classical fluid mechanics). 

The GP theory starts from the many--particle Hamiltonian of condensed matter physics in the second quantization \eqref{GP Hamiltonian}. When applied to a system of interacting bosons, it provides a connection to Bose-Einstein condensation (justifying the property of being a superfluid), and it leads to the existence of quantized vortices \cite{pitaevski-book,krstulovic2023}. On the other hand, the GP theory is a good picture for dilute quantum gases, where the contact interaction $\hat U=g \delta^{(3)}(\mathbf r - \mathbf r')$ is weak. In superfluid helium-4, the interactions between atoms are known to be strong \cite{tilley}. Moreover, it is typically too computationally demanding to solve the GP equation for large number of particles, and the GP theory is thus not suitable for macroscopic flows.

\subsection{Quantum Turbulence} \label{sec: quantum turbulence}
Instead of solving the GP equation, one can also assume that $\nabla\cdot\vv^s=0$ and obtain superfluid velocity from the vortex filaments by the Biot-Savart formula \cite{haenninen14},
\begin{equation} \label{BiotSavart vs}
    \mathbf v^s(\mathbf r,t)=\frac{\kappa}{4\pi} \int \frac{(\mathbf s- \mathbf r)\times d\mathbf s}{|\mathbf s -\mathbf r|^3}
\end{equation}
where $\mathbf s$ describes the position vector along a quantized vortex \cite{HänninenBaggaley14} and $\kappa$ is a quantum of circulation. This is the basis of the vortex filament method \cite{VFM}. The vortex filament method can be then coupled with hydrodynamic equations for the normal velocity, providing mutual coupling \cite{galantucci2020}. The vortex-filament method is, however, still quite computationally expensive for macroscopic systems. Moreover, it assumes that $\vv^s$ is divergence free. Therefore, we aim at hydrodynamic approaches toward the dynamics of He II, briefly recalled in the following section.

\subsection{Two-fluid models}\label{sec:HVBK}
The basic Landau--Tisza two--fluid model starts a phenomenological description of He II from what is known in classical hydrodynamics. The liquid is assumed to be composed of two kinds of densities, $\rho_n$ and $\rho_s$, and two kinds of momentum densities $\mm_s$ and $\mm_n$ \cite{tisza,Landau1941}. One part of the liquid is given by the superfluid component, $\rho_s \mathbf{v^s}$, whereas the other part is given by the normal component, $\rho_n \mathbf v_n$. 

However, the Landau--Tisza model was found to be too simple to describe other features of Helium II. The reason is that the model does not describe quantized vortices \cite{onsager1949,feynman1955,vinen1957,donnelly1999,bss}.
Quantized vortices are usually modeled as cylindrical tubes of length $\ell\gg a$, which can form rings or terminate on both ends on surfaces. From vortex-ring-propagation experiments \cite{rayfield}, it follows for their core diameter $a$ that $a \sim 10^{-10}$~m. Hence, their thickness lies in the length scale of atomic or quantum physics, whereas their longitudinal extent can be many orders larger and therefore in the mesoscopic or macroscopic length scale.
Quantized vortices evolve within the liquid, they mediate the interaction between the fluid components in two-fluid model via mutual friction, and attenuate the second sound \cite{HallVinen56, HallVinen56_2,Varga19,Novotny24}. 
It is, therefore, necessary to cover the physics of quantized vortices in models of superfluid helium-4. 


The standard generalization of the Landau--Tisza model that takes into account quantized vortices is the Hall-Vinen-Bekarevich-Khalatnikov (HVBK) model \cite{HallVinen56,BK1961}.
Let us recall set of HVBK equations in the form of \cite{donnelly1999}. The system of evolution equations consists of the assumed incompressibility conditions of the two components,
\begin{subequations}\label{eq.HVBK}
\begin{equation}
\nabla\cdot\vv^s = 0 = \nabla\cdot \vv_n,
\end{equation}
and two velocity equations,
\begin{align}
\partial_t \vv_n + \vv_n\cdot\nabla \vv_n &= -\nabla p_n +\nu_n \Delta \vv_n + \frac{\rho_s}{\rho}\FF_{ns}\\
\label{eq.HVBK.vs}\partial_t \vv^s + \vv^s\cdot\nabla \vv^s &= -\nabla p_s + \TT - \frac{\rho_n}{\rho}\FF_{ns}
\end{align}
supplied with constitutive relations for the respective pressures
\begin{equation}
\nabla p_n = \frac{1}{\rho} \nabla P + \frac{\rho_s}{\rho_n} \frac{s}{\rho} \nabla T
\quad \text{and}\quad
\nabla p_s = \frac{1}{\rho} \nabla P - \frac{s}{\rho} \nabla T,
\end{equation}
and, finally, the tension force $\TT$ and the mutual friction force $\FF_{ns}$,
\begin{align}
\TT &= -\nu_s \oomega \times (\nabla\times \hat{\oomega})\\
\FF_{ns} &= \frac{1}{2}B \hat{\oomega} \times (\oomega\times \cc) + \frac{1}{2}B' \oomega \times \cc
\end{align}
with
\begin{equation}
\cc = \vv_n - \vv^s - \nu_s \nabla\times \hat{\oomega} 
\quad\text{and}\quad
\nu_s = \frac{\kappa}{4\pi}\ln\left(\frac{R}{a}\right).
\end{equation}
\end{subequations}
In the absence of quantized vortices, the mutual friction force vanishes and the Landau--Tisza model is recovered. Superfluid vorticity, $\oomega=\nabla \times \mathbf v^s$, does not need to be zero, as the model is macroscopic and thus $\oomega$ stands for the average vorticity contained within a representative volume ($\hat \oomega=\oomega/|\oomega|$ being the direction of vorticity). Coefficient $\nu_s$ is the vortex tension parameter, $R$ is the mean distance between vortices \cite{Holm-HVBK}, $a$ is the vortex core radius, and coefficients $B$ and $B'$ are to be determined experimentally \cite{barenghi1983}. 

There are, however, many versions and modifications of the HVBK equations \cite{HallVinen56,BK1961,Holm-HVBK}. The two pressures contain contributions from the counterflow velocity, $p_n = \frac{\rho_n}{\rho}(p+\rho_s|\vv_n-\vv^s|^2)$ and $p_s = \frac{\rho_s}{\rho}(p-\rho_n |\vv_n-\vv^s|^2)$, while keeping both components incompressible \cite{boue2015}, \cite{nemirovskii2022}. A comprehensive review of superfluid dynamics was given in \cite{nemirovskii2013}. A way to better understand the origin of the possible variants is to formulate the model geometrically so that its form becomes invariant with respect to transformations of the state variables. We mainly use Hamiltonian mechanics as it is the geometry of analytical mechanics \cite{arnold,landau1,abraham-marsden} even in the case of complex fluids \cite{gaybalmaz-ratiu,pavelka-ijes,pkg}.

\subsection{Geometric and one-fluid models}\label{sec.geometric.models}
Hamiltonian models of superfluid Helium date back to the early works of Kronig and Thellung \cite{KronigThellung52}, Ziman \cite{Ziman53}, Pokrovskii and Khalatnikov \cite{Prokrovsii_Khalatnikov76}, Khalatnikov \cite{khalatnikov}, Volovick and Dzyaloshinskii \cite{dv}, or Gay-Balmaz and Ratiu \cite{gaybalmaz-ratiu}. Those works can be considered as a geometric justification of the quantum hydrodynamics introduced by Landau \cite{Landau1941} and provide a general Hamiltonian theory of superfluid helium. Due to the formal correspondence between the Poisson brackets and the commutator algebra in analytical mechanics and quantum mechanics, those Hamiltonian models can, in principle, be seen as rooted in quantum theory (see also Section \ref{subsec:correspondence}). A comprehensive geometrical derivation of HVBK equations with some auxiliary fields, based on the assumption of an underlying Lie algebra, was provided in \cite{Holm-HVBK,grmela-superfluid}. 

In the current manuscript, we introduce a geometric one-fluid model within the General Equation for Non-Equilibrium Reversible-Irreversible (GENERIC) framework \cite{go,og,hco,pkg}, which combines the Hamiltonian approach (reversible part) with generalized gradient dynamics (irreversible part), similarly as \cite{dv}, but we also go beyond the two-fluid model. In particular, our one-fluid model contains self-advection of the superfluid velocity in the presence of vorticity, and for high enough counter-flow velocities, normal density becomes a quadratic function of that velocity.

As helium-4 does not contain two kinds of atoms, one can not separate the superfluid with mass density $\rho_s$ from the normal fluid with $\rho_n$, which may be seen as a drawback of the two-fluid models. Instead, the liquid can be described as one fluid with two motions. In a series of works \cite{Mongiovi1991,mongiovi1993-stress, Mongiovi2017,Mongiovi2018}, Mongiov{\`i} and colleagues have shown that the two-fluid model can be reformulated as a one-fluid model with hyperbolic heat conduction \cite{catt} or with an evolution equation for the stress tensor within Extended Irreversible Thermodynamics \cite{jou-eit}. The geometric one-fluid model presented in this manuscript can be seen a further step in this direction. 

\section{A geometric one-fluid model}\label{sec.one}
Instead of working with two densities $\rho_n$ and $\rho_s$, we propose to work only with the total density $\rho$, as superfluid helium-4 consists of only one kind of atoms. Helium-4 can be then described by a one-fluid model based on the following assumptions:
\begin{itemize}
    \item The model should have microscopic origins.
    \item It should be geometrically consistent, that is, formulated within differential geometry.
    \item It should be close to the Landau--Tisza and HVBK two-fluid models and the vorticity of the superfluid velocity should be non-zero.
\end{itemize}
From the microscopic point of view, liquid helium-4 at rest behaves as a collection of excitations with a dispersion relation similar to phonons \cite{landau9,feynman1953} in the low-energy limit. For higher energies, the dependence of energy on the wave number ceases to be linear (roton and maxon parts of the energy spectrum) \cite{Godfrin2021}. Spatial variations of the excitations can be described by the distribution function $f^{ex}(\rr,\pp)$, where $\rr$ is the position and $\pp$ is the momentum \cite{peiersl}. This kinetic description, which we also use in this manuscript, is restricted to low temperatures  ($T<1.3K$), where the excitations can be treated as an ideal Bose gas \cite{landau5, landau9} and where the dispersion relation is approximately independent of temperature \cite{Godfrin2021}.
This microscopic picture is then to be superimposed to the macroscopic motion of the liquid.

Geometric consistency is provided by the GENERIC structure of the evolution equations, where the reversible part of the equations is generated by a Poisson bracket and energy (Hamiltonian) while the irreversible part is generated by a dissipation potential and entropy. We could also use the single-generator formalism \cite{be} or metriplectic framework \cite{mor,morrison4}. 

The evolution equation of the distribution function $f^{ex}(\rr,\pp)$, in particular, has a well-known Hamiltonian part \cite{grcontmath,mor}. The macroscopic motion can be also described by a distribution function of the average positions and momenta of the atoms (not taking into account the excitations), $f^m(\rr,\pp)$, from which the mass density $\rho$ can be obtained by averaging. This distribution function has also a Hamiltonian evolution. The division of the dynamics to excitations and macroscopic motion is similar to a crystal with excitations (phonons) that is not ideal but rather starts deforming and flowing. 

To formulate a model close to the two-fluid approaches, we shall work with hydrodynamic quantities instead of the distribution functions. The mass density is obtained from the atom distribution function $f^m(\rr,\pp)$ as $\rho = \int m f^m \dd\pp$, where $m$ is the mass of one helium-4 atom. The momentum density of the average mass motion $\mm^m$ is given by $\mm^m = \int \pp f^m \dd\pp$. The superfluid velocity $\vv^s$ is then defined as the average velocity of the atoms, $\vv^s = \mm^m/\rho$. As there is no superfluid component in our model, the role of superfluid velocity is played by the overall motion of atoms. The entropy density $s$, which is connected with transfer of heat by the excitations, is the entropy of bosons with distribution function $f^{ex}$ \cite{landau5}. Finally, as the excitations also carry momentum (despite being massless), the total momentum is the sum of the momentum of the macroscopic motion and the momentum of the excitations. The model then has the following state variables: the total density $\rho$, superfluid velocity $\vv^s$, total momentum density $\mm$, and entropy density $s$. Figure \ref{fig.model} shows the structure of the model. 
\begin{figure}
\centering
\begin{tikzpicture}
    \node (mass) at (0,3) {atoms};
    \node (top) at (0,2) {$(\rr,\pp)$};
    \node (middle) at (0,1) {$f^m(\rr,\pp)$};
    \node (bottom) at (0,-1) {$\begin{aligned}\rho &= \int m f^m \dd\pp \\ \mm^m &= \int \pp f^m \dd\pp \end{aligned}$};
    
    \draw[->] (top) -- (middle);
    \draw[->] (middle) -- (bottom);
    \draw[dashed] (2,3) -- (2,-2);
    
    \node (mass) at (4,3) {excitations};
    \node (top2) at (4,2) {$(\rr,\pp)$};
    \node (middle2) at (4,1) {$f^{ex}(\rr,\pp)$};
    \node (bottom2) at (4,-1) {$\begin{aligned}s &= \int \eta(f^{ex}) d\pp\\ \mm^{ex} &= \int \pp f^{ex} d\pp\end{aligned}$};
    
    \draw[->] (top2) -- (middle2);
    \draw[->] (middle2) -- (bottom2);

     \draw[dashed] (-1,-2) -- (5,-2);
    \node (transformation) at (2,-2.5) {superfluid helium};
    \node (transformation) at (2,-3) {$\rho, \mm = \mm^m+\mm^{ex}, s, \vv^s = \frac{\mm^m}{\rho}$};
\end{tikzpicture}
    \caption{A one-fluid model of superfluid helium-4. Motion of excitations is combined with the macroscopic motion (the average motion of atoms). The mass density then comes from the average positions of atoms, the momentum density is the sum of the momentum of the macroscopic motion and the momentum of the excitations, and the superfluid velocity is the average velocity of atoms. Entropy density is connected with the excitations and depends on their distribution function as in a Bose gas, see \cite{landau5} or Equation \eqref{eq.s.ph}.}
\label{fig.model}
\end{figure}

In the one-fluid model, vorticity of the superfluid velocity $\oomega = \nabla\times \vv^s$ does not need to be zero because the model works on a macroscopic scale, that is on scales larger than the size of the vortex core. The vorticity is then seen as the average vorticity contained in a representative volume.

In the following sections, we derive the Poisson bracket and energy that govern the reversible evolution of the model. Then, we add dissipation in order to describe also the irreversible evolution.

\subsection{Poisson brackets}
To derive the Poisson bracket generating the reversible evolution of the one-fluid model, we start with the Poisson brackets for the evolution of the underlying distribution functions. Then, those brackets are projected to the hydrodynamic state variables of the one-fluid model. When the Poisson bracket is known, the reversible part of evolution of the state variables can be obtained by the Hamiltonian equations of motion,
\begin{equation}
    (\partial_t \xx)_{rev} = \{\xx,E\},
\end{equation}
where $\xx$ are the state variables and $E$ is the energy (or Hamiltonian). 

The Poisson bracket $\{\bullet,\bullet\}$ is in general a bilinear operator that satisfies the Leibniz rule, antisymmetry, and the Jacobi identity, see for instance \cite{marsden-ratiu, hco, pkg}. From the physical point of view, bilinearity means that if we have two non-interacting systems, their evolution is the sum of the two respective evolutions. Leibniz rule makes the resulting evolution equations invariant with respect to constant energy shifts. Antisymmetry ensures that the energy is automatically conserved. Finally, Jacobi identity implies that the Poisson bracket (or more precisely, the underlying Poisson bivector \cite{fecko}) is invariant with respect to the evolution (its Lie derivative being zero) \cite{LagEul}.

\subsubsection{Kinetic theory of quasiparticles and matter}
The Poisson bracket for the distribution function $f^{ex}(t,\rr,\pp)$ of the quasiparticles (or excitations) and $f^m(t,\rr,\pp)$ of average atom positions and their momenta is given by the sum of their respective Poisson brackets \cite{grcontmath,mor},
\begin{align}\label{eq.PB.kinetic}
    \{F,G\}^{(\text{kinetic})} =& \int \drr\int d\pp f^{ex}(t,\rr,\pp) \left(\frac{\partial F_{f^{ex}}}{\partial \rr} \frac{\partial G_{f^{ex}}}{\partial \pp} -\frac{\partial G_{f^{ex}}}{\partial \rr} \frac{\partial F_{f^{ex}}}{\partial \pp}\right)\nonumber\\
    &+ \int \drr\int \dpp f^m(t,\rr,\pp) \left(\frac{\partial F_{f^m}}{\partial \rr} \frac{\partial G_{f^m}}{\partial \pp} -\frac{\partial G_{f^m}}{\partial \rr} \frac{\partial F_{f^m}}{\partial \pp}\right).
\end{align}
Note that subscripts following capital letters (functionals) stand for functional derivatives, for instance $F_{f^{ex}} \equiv \frac{\delta F}{\delta f^{ex}}$, throughout the manuscript. For functional derivatives of fields we do not use the subscript shorthand to avoid confusion.
Poisson bracket \eqref{eq.PB.kinetic} can be seen as a consequence of the validity of the canonical Poisson bracket for positions and momenta \cite{marsden-bbgky,gpd2}, and the canonical Poisson bracket can be seen as a consequence of the Correspondence Principle between quantum and classical mechanics, see Section \ref{subsec:correspondence}.
This bracket would generate the evolution equations of the quasiparticles and average atom positions similar to that in the kinetic theory \cite{struch}. However, our goal is a hydrodynamic model, so we have to project this Poisson bracket onto the hydrodynamic state variables.

\subsubsection{Hydrodynamic Poisson bracket}\label{sec.PB}
The kinetic Poisson bracket \eqref{eq.PB.kinetic} can be reduced to a hydrodynamical Poisson bracket by projecting the distribution functions onto the mass density, mass momentum, entropy density, and phonon momentum,
\begin{subequations}\label{eq.hydro}
    \begin{align}
        \rho &= \int \dpp m f^m,\\
        \mm^m &= \int \dpp \pp f^m\\
        s &= \int \dpp \eta(f^{ex}),\\
        \mm^{ex} &= \int \dpp \pp f^{ex}
    \end{align}
    where $\eta(f^{ex})$ is a smooth function of $f^{ex}$ (specified later in Equation \eqref{eq.s.ph}).
\end{subequations}
The assumption that the system is well described by these hydrodynamic fields actually means that our spatial scale is coarse-grained to some suitable representative volumes. Then, when integrating with respect to the space, we actually count over the small representative volumes $\drr$, which is a form of double averaging. 
When plugging functionals dependent on the distribution functions only through hydrodynamic fields \eqref{eq.hydro} into the Poisson bracket, the bracket then reduces to 
\begin{align}\label{eq.PB.hydrodynamic}
    \{F,G\}^{(\text{hydrodynamic})} =& 
    \int\drr m^{ex}_i \left(\partial_j F_{m^{ex}_i}G_{m^{ex}_j} -
\partial_j G_{m^{ex}_i}F_{m^{ex}_j}\right) 
+ \int\drr s \left(\partial_i F_s
G_{m^{ex}_i} - \partial_i F_s G_{m^{ex}_i}\right)\nonumber\\
&+ \int\drr m^m_i \left(\partial_j F_{m^m_i}G_{m^m_j} -
\partial_j G_{m^m_i}F_{m^m_j}\right) 
+ \int\drr \rho \left(\partial_i F_\rho
G_{m^m_i} - \partial_i F_\rho G_{m^m_i}\right).
\end{align}
The reduction works for any smooth function $\eta$, which is sometimes called a small miracle \cite{abarbanel}, and a detailed calculation can be found for instance in \cite{pkg}.
This Poisson bracket generates the reversible evolution equations of state variables \eqref{eq.hydro}. 
Note that again subscripts next to functionals denote functional derivatives, for instance $F_{m^{ex}_i} \equiv \frac{\delta F}{\delta m^{ex}_i}$.

To formulate a model of superfluid helium-4 similar to HVBK, we transform the state variables to 
\begin{equation}
(\rho, \mm^m, s, \mm^{ex}) \rightarrow (\rho, \mm = \mm^m+\mm^{ex}, s, \vv^s = \mm^m/\rho),
\end{equation}
where the superfluid velocity is defined as the ratio of the mass momentum and density. The hydrodynamic Poisson bracket then turns to the HVBK Poisson bracket
\begin{align}\label{eq.PB.HVBK}
    \{F,G\}^{(\text{HVBK})} &=\int\drr \rho\left(\partial_i F_\rho G_{m_i}-\partial_i G_\rho F_{m_i}\right)\nonumber\\
      &\quad+ \int\drr m_i\left(\partial_j F_{m_i} G_{m_j}-\partial_j G_{m_i} F_{m_j}\right)\nonumber\\
      &\quad+ \int\drr s\left(\partial_j F_{s} G_{m_j}-\partial_j G_{s} F_{m_j}\right)\nonumber\\
    &\quad+\int\drr \left(G_{v^s_{i}} \partial_i F_\rho - F_{v^s_{i}} \partial_i G_\rho\right) \nonumber\\
&\quad+ \int\drr v^s_{j}\left(\partial_i F_{v^s_{i}} G_{m_j} - \partial_i G_{v^s_{i}} F_{m_j}\right) \nonumber\\
&\quad+ \int\drr \left(\partial_i v^s_{j} -\partial_j v^s_{i}\right) \left(F_{v^s_{i}}G_{m_j}-G_{v^s_{i}}F_{m_j}\right)\nonumber\\
&\quad+ \int \drr \frac{1}{\rho}(\partial_i v^s_{j}-\partial_j v^s_{i}) F_{v^s_{i}}G_{v^s_{j}}.
\end{align}
This bracket has been found also in \cite{pof2021} as the unique solution of Jacobi identity and in \cite{volovick-dotsenko1980} by explicitly taking into account the vortex filaments.

Poisson bracket \eqref{eq.PB.HVBK} generates the reversible evolution equations
\begin{subequations}\label{eq.evo.MaHe.rev}
\begin{eqnarray}
    \label{eq.rho.evorev}\left(\frac{\partial \rho}{\partial t}\right)_{rev} &=& -\partial_k (\rho E_{m_k}+E_{v^s_{k}})\\
    \label{eq.m.evorev}\left(\frac{\partial m_i}{\partial t}\right)_{rev} &=& 
-\partial_j(m_i E_{m_j}) -\partial_j(v^s_{i} E_{v^s_{j}}) - \rho\partial_i E_\rho -m_j \partial_i E_{m_j} - s\partial_i E_s -v^s_{k} \partial_i E_{v^s_{k}}+\partial_i(E_{v^s_{k}} v^s_{k})\\
 \label{eq.s.evorev}\left(\frac{\partial s}{\partial t}\right)_{rev} &=& -\partial_k \left(s E_{m_k}\right)\\
    \label{eq.vs.evorev}\left(\frac{\partial v^s_{k}}{\partial t}\right)_{rev} &=& -\partial_k E_\rho -\partial_k(v^s_{j}  E_{m_j}) 
+(\partial_k v^s_{j} - \partial_j v^s_{k})\left(E_{m_j}+\frac{1}{\rho}E_{v^s_{j}}\right).
\end{eqnarray}
These equations will be combined with the irreversible evolution in Section \ref{sec.dissipation}.
\end{subequations}

The stress tensor can be identified as the terms under the divergence in the momentum equation \eqref{eq.m.evorev},
\begin{equation}
T_i^{\,\,j}=-\left(\bar{P} - \vv^s\cdot \frac{\partial e}{\partial \vv^s}\right)\delta_i^j - m_i \frac{\partial e}{\partial m_j} - v^s_i \frac{\partial e}{\partial v^s_j}
+v^s_i \epsilon^{jlm}\partial_l \frac{\partial e}{\partial \omega^m} - \epsilon^{jmk}\frac{\partial e}{\partial \omega^m} \partial_i v^s_k
\end{equation}
where the generalized pressure $\bar{P}$ is defined as the complete Legendre transformation of the volumetric energy density $e$,
\begin{equation}
    \bar{P} = -e + \rho \frac{\partial e}{\partial \rho} + s \frac{\partial e}{\partial s} + \mm\cdot \frac{\partial e}{\partial \mm} + \vv^s\cdot \frac{\partial e}{\partial \vv^s},
\end{equation} 
where the volumetric energy density $e(\rho,\mm,s,\vv^s,\oomega)$ is a function of the state variables and $\oomega$. The total energy can be obtained by integrating the volumetric energy density over the whole volume, $E = \int\drr e$.
Total momentum is conserved, as the equation for the momentum density is in the divergence form.

Note that if the superfluid vorticity $\oomega = \nabla\times\vv^s$ is zero initially, the reversible evolution keeps it zero. Although superfluid vorticity is believed to be zero at the microscopic level (in the Gross-Pitaevskii model, for instance), the one-fluid model works on a macroscopic scale where $\oomega$ may be seen as an average over the volume containing many vortex filaments. Therefore, it does not need to be zero in general.

Evolution equations \eqref{eq.evo.MaHe.rev} yield also the reversible part of the evolution equation for the volumetric energy density, 
\begin{align}\label{eq.Ediv}
   (\partial_t e)_{rev} &= -\nabla\cdot\left(\left(\rho E_\rho  +s E_s + \mm\cdot E_\mm\right)\mm +E_\rho E_{\vv^s} + (\vv^s\cdot E_\mm) E_{\vv^s}\right)\\
   &= -\nabla\cdot\left(\left(e+\bar{P}+\vv^s\cdot\left(\nabla\times\frac{\partial e}{\partial \oomega}\right)\right)\frac{\partial e}{\partial \mm} + \left(\left(\vv^s\cdot \frac{\partial e}{\partial \mm} + \frac{\partial e}{\partial \rho}\right)\mathbb{I}-\frac{\partial e}{\partial \mm}\otimes\vv^s\right)\cdot \left(\frac{\partial e}{\partial \vv^s} + \nabla\times \frac{\partial e}{\partial \oomega}\right)\right), \nonumber
\end{align}
using the chain rule. Note that $\mathbb{I}$ stands for the unit tensor. Equation \eqref{eq.Ediv} is in the divergence form and the total energy is thus conserved.

\subsubsection{Clebsch formulation}
Poisson bracket \eqref{eq.PB.hydrodynamic} can be also seen from the perspective of the Clebsch formulation \cite{clebsch,pkg}, which brings to the Hamiltonian evolution a variational formulation \cite{gaybalmaz-open}. A Clebsch formulation has the symplectic structure, which is indeed directly related with an Euler-Lagrange equation \cite{fecko}, and a variational formulation is often a desired feature of evolution equations. 

Within the Clebsch formulation, pairs of state variables are equipped with the canonical Poisson bracket that comes right from the principle of stationary action. In fluid mechanics, the Clebsch variables can be chosen as pairs ($\rho$, $\rho^*$), ($s$, $s^*$), and ($\alpha$, $\alpha^*$). In the case of superfluid helium-4, we add one more pair ($\beta$, $\beta^*$), which will allow for the superfluid vorticity. The canonical Poisson bracket for the Clebsch variables is given by
\begin{multline}
    \{F,G\}^{Clebsch} = \int \drr \left(F_\rho G_{\rho^*}-G_\rho F_{\rho^*}\right)
    +\int \drr \left(F_s G_{s^*}-G_s F_{s^*}\right)\\
    +\int \drr \left(F_\alpha G_{\alpha^*}-G_\alpha F_{\alpha^*}\right)
    +\int \drr \left(F_\beta G_{\beta^*}-G_\beta F_{\beta^*}\right),
\end{multline}
where, similarly as before, subscripts following functionals denote functional derivatives, for instance $F_\rho \equiv \frac{\delta F}{\delta \rho}$.
Then, the state variables of the one-fluid model can be expressed in terms of the Clebsch variables as
\begin{equation}
    \rho = \rho, s = s, \mm^m = \rho\nabla\rho^* + \alpha\nabla\alpha^*, \mm^{ex} = s\nabla s^* + \beta\nabla\beta^*,
\end{equation}
and the Clebsch Poisson bracket reduces to the hydrodynamic Poisson bracket \eqref{eq.PB.hydrodynamic}, similarly as in fluid mechanics \cite{bedeaux-clebsch,pkg}.

Let us now add a few remarks to build intuition about the Clebsch variables, starting from classical fluid mechanics. First, the fields $\rho$ and $s$ are the standard mass density and entropy density (per volume). If we had only the pair $(\rho,\rho^*)$, the momentum density would be $\mm^{(\rho)} = \rho\nabla\rho^*$, and the Poisson bracket for $\rho$ and $\mm^{(\rho)}$ would be the fluid-mechanics bracket for density and momentum density,
$$\int\drr \rho\left(\partial_i F_\rho G_{m_i^{(\rho)}} - \partial_i G_\rho F_{m_i^{(\rho)}}\right)
+ \int\drr m_i^{(\rho)}\left(\partial_j F_{m_i^{(\rho)}} G_{m_j^{(\rho)}} - \partial_j G_{m_i^{(\rho)}} F_{m_j^{(\rho)}}\right),$$
which is the standard Poisson bracket for barotropic fluids \cite{arnold}.
For a Hamiltonian with the standard kinetic energy density $\frac{(\mm^{(\rho)})^2}{2\rho}$, velocity, which is the derivative of the Hamiltonian with respect to the momentum, would be curl free, $\vv =\frac{\mm^{(\rho)}}{\rho} = \nabla\rho^*$. The Clebsch variable $\rho^*$ can be thus interpreted as the velocity potential. In the context of superfluid helium-4, this potential is proportional to the phase of the wave-function. 

If we now add the pair $(\alpha,\alpha^*)$, the momentum density becomes $\mm^{(\rho,\alpha)} = \rho\nabla\rho^* + \alpha\nabla\alpha^*$, while keeping the same Poisson bracket for $\rho$ and $\mm^{(\rho,\alpha)}$. The velocity now no longer needs to be curl-free, $\vv = \nabla\rho^* + \frac{\alpha}{\rho}\nabla\alpha^*$, and thus the pair $(\alpha,\alpha^*)$ brings vorticity into the dynamics.

If we now add the pair $(s,s^*)$, we obtain the standard momentum density in fluid mechanics, $\rho\nabla\rho^* + \alpha\nabla\alpha^* + s\nabla s^*$, and the standard Poisson bracket of fluid mechanics \cite{dv,mor,grpd,pkg}. Note, however, that the number of variables in fluid mechanics is five (mass density, entropy density, and three components of the momentum density), while the number of state variables in the corresponding Clebsch formulation is six (three pairs). The step from the Clebsch formulation to fluid mechanics can be thus seen as a reduction.

In the geometric one-fluid model of superfluid helium-4, we form a fluid-mechanics Poisson bracket from the pairs $(\rho,\rho^*)$ and $(\alpha,\alpha^*)$, which brings the dynamics of density and superfluid velocity. The other two pairs, ($s,s^*$) and $(\beta,\beta^*)$, are used to bring the dynamics of entropy density and momentum density of excitations (related to the heat flux). If, for instance, we discarded the pair $(\alpha,\alpha^*)$, the superfluid velocity $\vv^s = \mm^{m}/\rho$ would be curl-free. Similarly, if we discarded the pair $(\beta,\beta^*)$, the momentum density of excitations would be curl-free, in contrast with the theoretical predictions and experimental observations of phonon vortices \cite{fononi,bao-vortices}.

If we were to formulate a variational principle for the one-fluid model, we might take the Hamiltonian in terms of the Clebsch variables and get the Lagrangian as the Legendre transformation. In this transformation, the conjugate variables would be mapped to the rates of the remaining variables, for instance $\rho^*\rightarrow \partial_t \rho$, etc. 

In contrast with \cite{Prokrovsii_Khalatnikov76}, where the pair of Clebsch variables ($\alpha$, $\alpha^*$) was missing, the superfluid velocity $\vv^s = \frac{\mm^m}{\rho}$ does not need to be curl-free in our model. If we chose such a simpler Clebsch representation, the superfluid velocity would curl-free and field $\rho^*$ could be interpreted as proportional to the phase of the wave function. The pair $(\alpha, \alpha^*)$ thus brings the freedom to add vortices to $\vv^s$.

In summary, the Clebsch formulation provides another formulation of the reversible part of the geometric one-fluid model, as well as a variational formulation. 

\subsection{Energy (Hamiltonian) in the one-fluid model}
While Poisson bracket \eqref{eq.PB.HVBK} generates the form of the reversible evolution in the one-fluid model, the energy is needed to close the evolution equations. The energy functional for the one-fluid model is derived here similarly as in Landau's work \cite{Landau1941} or \cite{dv}. The energy of excitations is taken as the energy of helium-4 at rest and, subsequently, the total energy is obtained by Galilean transformation to the laboratory frame, in which the matter is moving. Let us start from the Galilean transformation as it gives the form of the energy functional.

\subsubsection{Galilean transformation to the laboratory frame}
Assume that the energy of helium-4 at rest is given by a function 
\begin{equation}\label{eq.ep}
    e^p = \frac{|\mm^{ex}|^2}{2\rho_n(\rho,s)} + e_0(\rho,s), 
\end{equation}
where $\mm^{ex}$ is the momentum density of excitations and where the internal energy $e_0$ is to be specified later. The energy of helium-4 in motion is then obtained by the Galilean transformation of the rest energy of helium-4 to the laboratory frame in which the liquid with mass density $\rho$ is moving with velocity $\vv^s$, 
\begin{equation} \label{energy functional}
E = \int \drr \left(\frac{1}{2}\rho |\vv^s|^2 + (\mm-\rho \vv^s)\cdot\vv^s + \frac{|\mm-\rho\vv^s|^2}{2\rho_n(\rho,s)}+e_0(\rho,s)\right),
\end{equation}
where the formula $\mm = \mm^{ex}+\rho \vv^s$ was used.
Note that $\rho_n(\rho,s)$ and $e_0(\rho,s)$ are two functions of $\rho$ and $s$. The normal density $\rho_n$ is not a state variable here. Instead, $\rho_n$ is a function of $\rho$ and $s$ that describes the effective inertia or the excitations (despite having no mass), see for instance \cite{fountain}. Also, as $\rho_n$ can be seen as a metric-like tensor, that provides the scalar product in $|\mm^{ex}|^2$, and in more complicated cases, $\rho_n$ can even be anisotropic \cite{volovik-droplet}. The rest energy of helium-4 is  addressed later by means of statistical mechanics in Sections \ref{sec.e.phonons} and \ref{sec.e.rotons}, where is shown that formula \eqref{eq.ep} is valid only approximately. 

Derivatives of energy \eqref{energy functional} with respect to the state variables are
\begin{subequations}
\begin{align}
\label{eq.dEdm}\frac{\delta E}{\delta \mm} =& \vv^s + \frac{\mm-\rho\vv^s}{\rho_n} \stackrel{def}{=}\vv_n\\
\frac{\delta E}{\delta \rho} =& -\frac{1}{2}|\vv^s|^2 - \frac{(\mm-\rho \vv^s)\cdot\vv^s}{\rho_n} - \frac{|\mm-\rho\vv^s|^2}{2\rho^2_n(\rho,s)}\frac{\partial \rho_n}{\partial \rho} + \underbrace{\frac{\partial e_0}{\partial \rho}}_{\stackrel{def}{=}\mu}\nonumber\\
 =& -\frac{1}{2}|\vv^s|^2 - \vv_{ns}\cdot\vv^s - \frac{1}{2}|\vv_{ns}|^2\frac{\partial \rho_n}{\partial \rho} + \mu(\rho,s)\\
\frac{\delta E}{\delta \vv^s} =& \mm - \rho \vv^s -\frac{\rho}{\rho_n}(\mm-\rho\vv^s) = -\rho_s\underbrace{(^\flat\vv_n-\vv^s)}_{\stackrel{def}{=}\vv_{ns}}\\
\frac{\delta E}{\delta s} =& - \frac{|\mm-\rho\vv^s|^2}{2\rho^2_n(\rho,s)}\frac{\partial \rho_n}{\partial s} + \underbrace{\frac{\partial e_0}{\partial s}}_{\stackrel{def}{=}T}
 = - \frac{1}{2}|\vv_{ns}|^2\frac{\partial \rho_n}{\partial s} +T,
\end{align}
where the normal velocity was defined as the derivative of the total energy with respect to the total momentum density. The superfluid density is defined as a function of $\rho$ and $s$, namely $\rho_s = \rho-\rho_n(\rho,s)$. Temperature $T$ and chemical potential $\mu$ are defined as the derivatives of $e_0$ with respect to $s$ and $\rho$, respectively. Note that neither $\rho_s$ nor $\vv_n$ play the role of state variables in the one-fluid model. Instead, they are functions of the state variables $\rho$, $\mm$, $s$, and $\vv^s$.
\end{subequations}

Equation \eqref{eq.dEdm} gives that 
\begin{equation}\label{eq.m.vv}
    \mm = \rho_s\vv^s+\rho_n {^\flat}\vv_n, 
\end{equation}
where $^\flat\vv_n$ denotes the covariant normal velocity, $(^\flat\vv_n)_i = g_{ij}v^j_n$ (the metric tensor is tacitly present in the scalar product $|\mm-\rho \vv^s|^2$). The counter-flow velocity is then defined as the difference between $^\flat\vv_{n}$ and $\vv^s$, that is $\vv_{ns} = ^\flat\vv_n-\vv^s$. Relation \eqref{eq.m.vv} makes the one-fluid model compatible with the two-fluid models, where the momentum of the normal fluid is $\rho_n\vv_n$ \cite{khalatnikov}. However, in the following sections, we show that the normal velocity is no longer linear in the momentum density, in contrast with the two-fluid models.

\subsubsection{Energy of phonons}\label{sec.e.phonons}
The energy of excitations in superfluid helium-4 is usually split into two contributions, one from phonons and one from rotons \cite{landau9}. 
The dependence of energy of helium-4 excitations on the $k-$vector in the region of small momenta is similar linear, similarly as that of phonons \cite{Godfrin2021}. For higher momenta, a local minimum in the energy appears and gives the roton contribution addressed in the following Section. Within these two Sections \ref{sec.e.phonons} and \ref{sec.e.rotons}, we derive in the phonon-roton splitting the classical formulas for free energy and normal density of Helium II, but we also obtain new contributions of higher order in $\vv_n$. Note that within this Section, we assume that $\vv^s = 0$, so the final dependence of the energy on $\vv_n$ has to be shifted.

Let us start with the phonon part. For phonons in the first Brillouine zone, where the dispersion relation is linear, we use the MaxEnt principle to derive the thermodynamic relations, similarly as \cite{anile,struch-dreyer,Larecki2010,fononi}. The energy-and-momentum densities
\begin{align}
    e^{ph} =& \int c |\pp| f^{ex}(\rr,\pp)\dd\pp\\
    \mm^{ph} =& \int \pp f^{ex}(\rr,\pp)\dd\pp,
\end{align}
where $c$ is the speed of sound.
When the distribution function of bosons is known, as in the kinetic theory, the volumetric entropy density is
\begin{equation}\label{eq.s.ph}
    s^{\text{kinetic\, theory}} = -\frac{k_B}{h^3}\int \dpp \left(h^3 f^{ex} \ln\left(h^3 f^{ex}\right) -\left(1+h^3 f^{ex}\right)\ln\left(1+h^3f^{ex}\right)\right),
\end{equation}
see \cite{landau5}.
The hydrodynamic entropy of phonons can be then calculated by means of the Maximum Entropy Principle (MaxEnt \cite{jaynes,fononi}), that leads to the distribution function
\begin{equation}\label{eq.f}
    f^{ph} = \frac{1}{h^3}\frac{1}{e^{\frac{\mm^*\cdot\pp}{k_B}}e^{\frac{e^*}{k_B}\varepsilon^{ph}(|\pp|)}-1}
\end{equation}
with $\varepsilon^{ph}(|\pp|) = c |\pp|$. Conjugate variables $\mm^*$ and $e^*$ are the derivatives of the total entropy with respect to the momentum density and total energy density, respectively, and they play the role of Lagrange multipliers within MaxEnt \cite{redext}.

After calculating the momentum density, see for instance \cite{fononi}, we can divide it by the normal velocity to obtain the normal density of phonons,
\begin{equation}
    \rho_n^{ph} = \frac{2\pi^2 (k_B T)^4 \left(1 + 3 \frac{|\vv_n|^2}{c^2}+ \mathcal{O}\left(\frac{|\vv_n|}{c}\right)^4\right)}{45 c^5 \hbar^3}.
\end{equation}
The normal density contains a contribution obtained by Landau, that is independent of $\vv_n$, and an additional contribution quadratic in $\vv_n$. Note, however, that the dependence on the normal velocity is restricted to the frame in which the superfluid velocity is zero. After the transformation to the laboratory frame, the normal density becomes a function of $\mm-\rho\vv^s$, which is Galilean invariant, see Section \ref{sec.beyond}.

The hydrodynamic phonon entropy is then
\begin{equation}\label{eq.sp}
    s^{ph}(\rr) = \frac{4}{3}\left(\frac{2\sigma}{c}\right)^{1/4}\left(e^{ph}(\rr)\right)^{3/4} \frac{3^{3/4}}{2^{7/4}}(3-\chi)^{1/2}(1-\chi)^{1/4}
\end{equation}
with Eddington factors $\chi = \frac{5}{3}-\frac{4}{3}\sqrt{1-\frac{3}{4}\frac{c^2 |\mm^{ph}(\rr)|^2}{e^{ph}(\rr)}}$, see \cite{struch-dreyer,fononi}.

The energy of phonons may be in principle obtained by inverting relation \eqref{eq.sp}, but not in the analytical form. Instead, we can use the approximate relation for the energy of phonons in the region of small momenta, expanding the entropy to the second order in the phonon momentum density, and obtain that 
\begin{align}\label{eq.e.ph}
    e^{ph} \approx& \epsilon + \frac{|\mm^{ph}|^2}{2\rho^{ph}_{n0}}\\
    \epsilon =& \frac{3^{4/3}}{8}\left(\frac{c}{\sigma}\right)^{1/3} \left(s^{ph}\right)^{4/3} = \frac{6\sigma}{c}T^4\\
    \rho^{ph}_{n0} =& \frac{4}{3}\frac{\epsilon(s)}{c^2} \approx \frac{8}{3}\frac{\sigma T^4}{c^3}
\end{align}
where $\sigma = \frac{\pi^2 k_B^4}{60 c^2 \hbar^3}$ is the Stefan-Boltzmann constant for the phonon sound speed $c$.
This gives an approximate relation for the phonon part of the normal fluid density function $\rho_n(s)$ and is in agreement with Landau's results \cite{landau9}. 

However, if we did not use the quadratic approximation, the dependence of energy on the phonon momentum would cease to be quadratic, which would violate assumption \eqref{eq.ep}, that was necessary for two-fluid models. This is even more important for the roton part of the spectrum, where the energy is not quadratic in the momentum, see Section \ref{sec.Godfrin}. The energy of rotons is calculated in the next Section.

\subsubsection{Energy of rotons}\label{sec.e.rotons}
The spectrum of helium-4 in a region of higher momenta is no longer linear and has a local minimum. This is the so called roton region and together with the phonon region, studied in the preceding Section, it gives a good approximation of the overall free energy. In this Section, we derive both the standard results and new a contribution of higher order in $\vv_n$. 

\begin{figure}[ht!]
    \centering
    \includegraphics[scale=0.5]{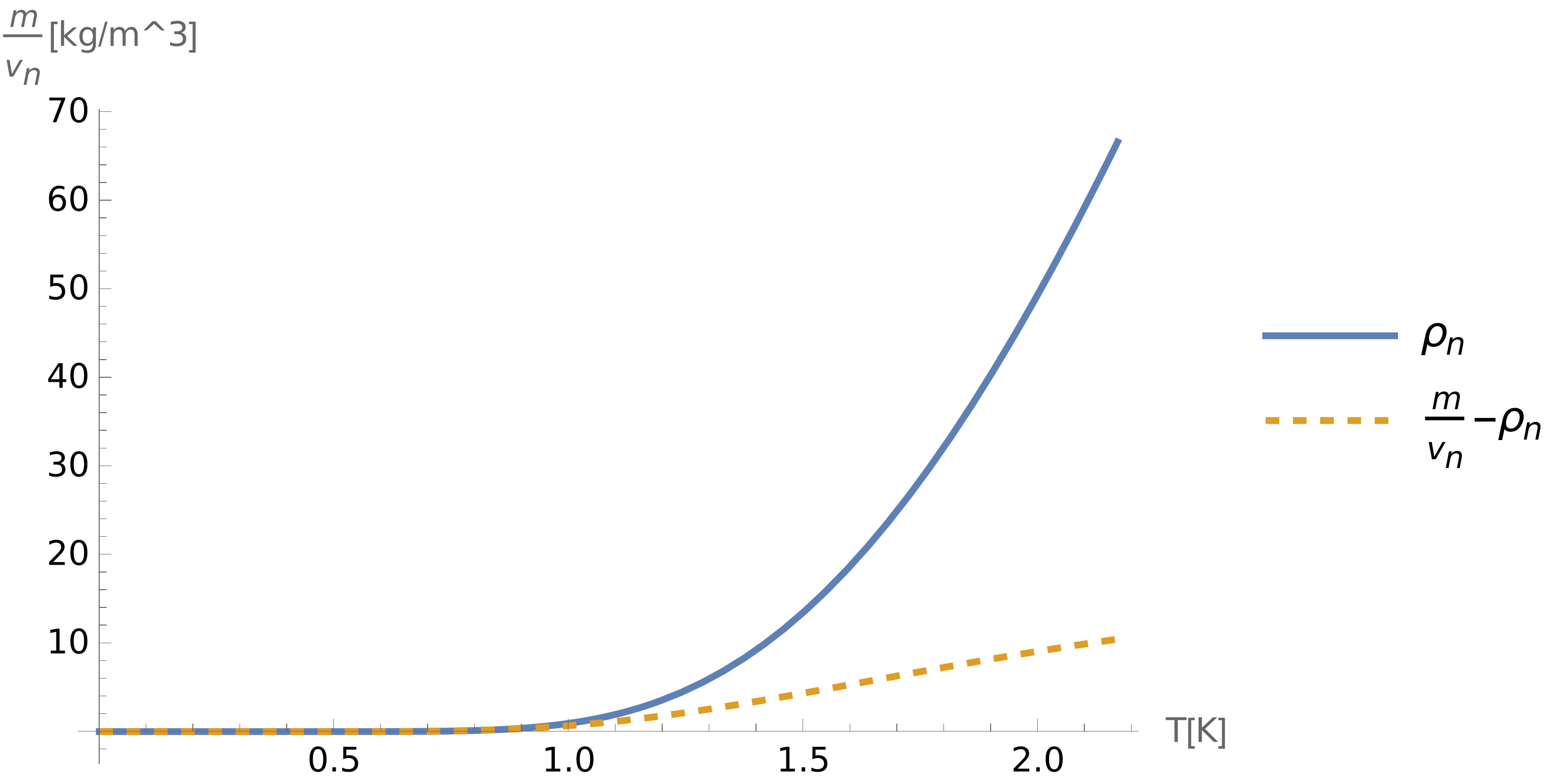}
    \caption{\label{fig.rhon}The dependence of $\rho^{rot}_n$ on temperature compared with the terms in momentum density that are of the third order in $\vv_n$ (for the particular value $\vv_n=18m/s$). The roton momentum density is only approximately proportional to $\vv_n$, similarly as the phonon momentum density is only approximately proportional to $\vv_n$. The two-fluid model is thus valid only approximately.}
\end{figure}

The energy density within the roton region can be approximated by
\begin{equation}
    e^{rot} = \int \dpp \left(\Delta + \frac{(|\pp|-p_0)^2}{2\mu_r}\right) f^{rot}(\rr,\pp)
\end{equation}
where $\Delta = 0.7418$~meV is the energy gap, $p_0 = 1.918\cdot 10^{10}\hbar$ is the momentum of the roton minimum, and $\mu_r = 0.141 m$ is the roton effective mass ($m$ being the mass of one helium atom) \cite{Landau1941,Godfrin2021}. 
For the roton part of the dispersion relation, the MaxEnt principle leads to the distribution function
\begin{equation}
    f^{rot} = \frac{1}{h^3}\frac{1}{e^{\frac{\mm^*\cdot\pp}{k_B}}e^{\frac{e^*}{k_B}\left(\Delta + \frac{(|\pp|-p_0)^2}{2\mu_r}\right)}-1}.
\end{equation}
The momentum density $\mm^{rot} = \int \dpp \pp f^{rot}$ is then approximately
\begin{equation}\label{eq.mrot}
    \mm^{rot} = \underbrace{\rho_{n0}^{rot} \left(1+\frac{\mu_r |\vv_n|^2}{5k_B T}\zeta F(\zeta)   + \mathcal{O}\left(\frac{\mu_r}{2k_B T}|\vv_n|^2\right)^{2}\right)}_{\stackrel{def}{=}\rho_n^{rot}(T,\vv_n)}\vv_n 
\end{equation}
where $\zeta(T) = \frac{p_0^2}{2\mu_r k_B T}$ is the dimensionless typical roton kinetic energy ($\approx 122$ at $T=1.3K$ and growing for smaller temperatures) and 
    $F(\zeta) = \frac{1+\frac{15}{2}\zeta^{-1}+\frac{45}{4}\zeta^{-2}+\frac{15}{8}\zeta^{-3}}{1+3\zeta^{-1}+\frac{3}{4}\zeta^{-2}}$.
Function $F$ varies linearly with temperature ($T\in [0,1.3]K$) going from $1.0$ to $1.045$, so it is close to one.
The static roton normal density$\rho_{n0}^{rot}(T) = \rho_n^{rot}|_{\vv_n=0}$, which represents the proportionality factor between $\mm^{rot}$ and $\vv_n$ in the limit of $\vv_n\rightarrow 0$, is 
\begin{align}\label{eq.rhon0}
    \rho_{n0}^{rot} =& e^{-\Delta/(k_B T)}\frac{\mu_r}{3(2\pi)^{3/2}}\left(\frac{k_B T \mu_r}{\hbar^2}\right)^{3/2}\left(3+12\frac{p_0^2}{2\mu_r k_B T} + 4\left(\frac{p_0^2}{2\mu_r k_B T}\right)^2\right)
    \left(1+\mathrm{Erf}\left(\frac{p_0}{\sqrt{2 k_B T \mu_r}}\right)\right)\nonumber\\
    \approx& e^{-\Delta/(k_B T)}\frac{8 \mu_r}{3(2\pi)^{3/2}}\left(\frac{k_B T \mu_r}{\hbar^2}\right)^{3/2}\left(\frac{p_0^2}{2\mu_r k_B T}\right)^2,
\end{align}
the approximated version of which is the standard Landau's formula for $\rho_n^{rot}$ \cite{landau9}, noting that the term with the $p_0^4$ is larger than the remaining terms by two orders of magnitude and that the $\mathrm{Erf}$ term is practically equal to unity. The dependence of $\rho_n^{rot}$ on the velocity is visible in Equation \eqref{eq.mrot} and momentum is thus not proportional to the normal velocity, similarly as in the case of phonons. The usual proportionality between $\mm$ and $\vv_n$ proposed by Landau is thus valid only approximately, as well as the two-fluid interpretation of superfluid helium-4, see Figure \ref{fig.rhon}.
Note also, that the normal density is not actually a function of $\vv_n$ but rather of the Galilean-invariant term $\mm-\rho\vv^s$, as the former is valid in the frame without superfluid velocity while the latter being in the laboratory frame, see Section \ref{sec.beyond}. 

Similarly, the energy becomes
\begin{align}
    e^{rot} \approx \frac{p_0^2 (k_B T+2\Delta)\sqrt{k_B T \mu_r}}{2\sqrt{2}\hbar^3 \pi^{3/2}}e^{-\Delta/k_B T} + \frac{|\mm^{rot}|^2}{2\rho_{n0}^{rot}(T)}
\end{align}
where we neglected the higher-order terms in $\vv_n$ and $\mm$. The entropy density of rotons is then
\begin{equation}
    s^{rot} = \frac{2 \sqrt{\mu_r T}p_0^2}{(2\pi)^{3/2} \hbar^3 }e^{-\Delta/k_B T}\left(\frac{3}{2}+\frac{\Delta}{k_B T}\right).
\end{equation}

Note that the calculated $\rho_n^{rot}$ and $\rho_n^{ph}$ give a good approximation of the experimental $\rho_n$ only up to the temperature of $1.3 K$. For instance, the total normal density $\rho_n = \rho_n^{ph}+\rho_n^{rot}$ at $T=2.17K$ gives approximately $0.065 g/cm^3$ instead of $0.145 g/cm^3$. The direct numerical integration of the dispersion relations without approximately splitting the integration to the phonon and roton parts gives a better agreement with the experimental data \cite{Godfrin2021}, but the result is sensitive on the high-energy end of the spectrum, and no results are provided for higher magnitudes of the wave-vector. To get a better and more reliable prediction by means of statistical mechanics, one would need to take care also of the high-energy part of the spectrum, as well as the interactions between the excitations, which is out of the scope of this paper.

Under the phonon-roton splitting approximation, the total energy density of the excitations is the sum of the phonon and roton contributions $e^p = e^{ph}+e^{rot}$, similarly as the momentum of the excitations is $\mm^{ex} = \mm^{ph}+\mm^{rot}$, and total entropy density $s = s^{ph}+s^{rot}$. The calculations of the phonon and roton quantities are done as if there were two independent excitation Bose gases. One should not however treat the phonon and roton contributions as separate state variables and only the total momentum density and total entropy density (or equivalent choices) should eventually be used. The number of variables is consistent by having the phonon and roton temperatures equal to each other, $\frac{\partial e^{ph}}{\partial s^{ph}} = T = \frac{\partial e^{rot}}{\partial s^{rot}}$, as well as the velocities $\frac{\partial e^{ph}}{\partial \mm^{ph}} = \vv_n = \frac{\partial e^{rot}}{\partial \mm^{rot}}$. The fundamental thermodynamic relation $e^p(\mm,s)$ should be then determined as the Legendre transformation of the overall free energy $f^{ex}(T,\vv_n) = f^{ph}(T,\vv_n)+f^{rot}(T,\vv_n)$, which would, however, have to be identified numerically, which is the purpose of the next Section.

\subsubsection{Energy based on the experimental dispersion relation}\label{sec.Godfrin}
Instead of using the approximate phonon and roton dispersion relations, one can use the experimental dispersion relation $\varepsilon(|\pp|)$ of helium-4 \cite{Godfrin2021} to calculate the energy of excitations numerically. Note that within this section, we assume that $\vv^s=0$, as the $\vv^s$ enters the Hamiltonian after the Galilean transformation. Note also that we use the experimental data for the saturated vapor pressure and to obtain also the dependence on the density, we would need to compile also data for different pressures, which is however, out of the scope of this paper.

Let us commence with the momentum density of the excitations, which is given by the integral 
\begin{equation}
    \mm^{ex} = \int  \pp f^{KT} \dd\pp,
\end{equation}
where the experimental $\varepsilon(|\pp|)$ is used in the distribution function. As a direct numerical integration would be challenging, we first rotate the coordinate system so that the conjugate momentum becomes $(0,0,|\mm^*|)$, and then we use the spherical coordinates in the momentum space and expand into series, keeping 
\begin{equation}
    |-\mm^*/e^*|=|\vv_n| < \frac{\varepsilon(|\pp|)}{|\pp|}\stackrel{def}{=}v_c.
\end{equation}
Note that the right-hand side of this inequality is the Landau's critical velocity ($v_c = 57.9m/s$ for the data from \cite{Godfrin2021}).

\begin{figure}[ht!]
    \centering
    \includegraphics[scale=0.5]{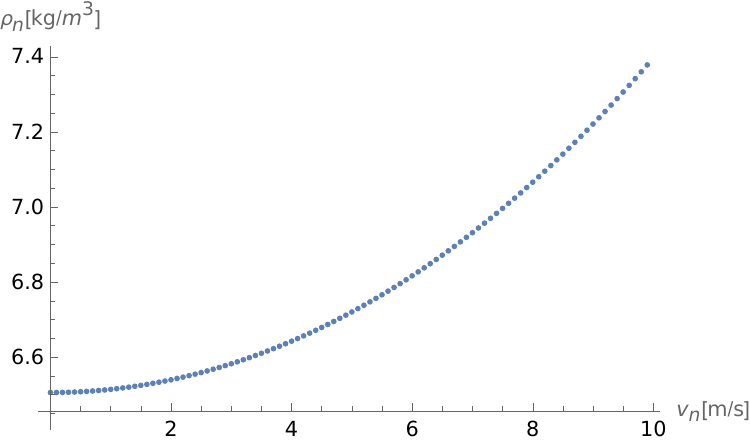}
    \caption{\label{fig.rhonT13}The dependence of the normal density $\rho_n^{ex}$ on the normal velocity $\vv_n$ at $T=1.3K$. For velocities higher than $5m/s$, the quadratic dependence of $\rho_n$ on $\vv_n$ becomes significant.}
\end{figure}

\begin{figure}[ht!]
    \centering
    \includegraphics[scale=0.5]{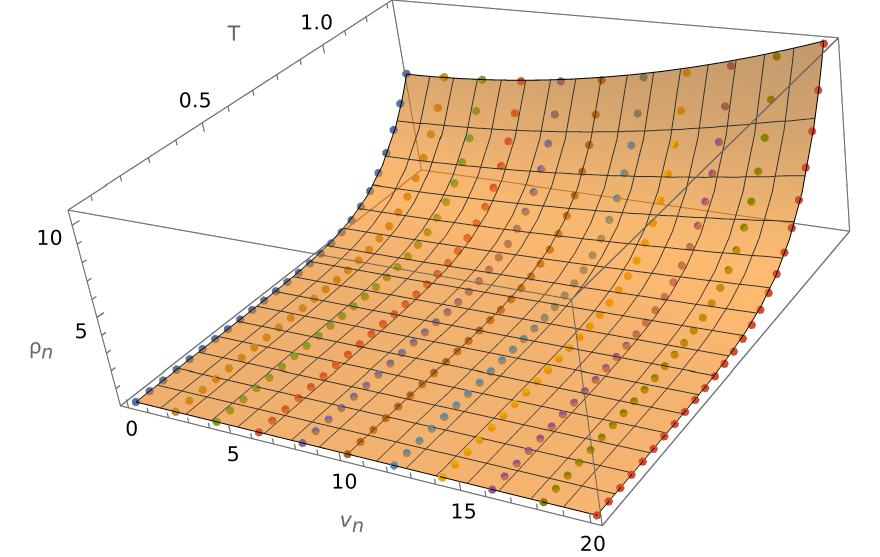}
    \caption{\label{fig.rhonfull}The dependence of the fitted normal density $\rho_n$ on temperature $T$ and velocity $\vv_n$. The normal density is approximately proportional to $T^7$ and contains a term quadratic in $\vv_n$. The numerical data (dots) are obtained from the experimental dispersion relation \cite{Godfrin2021}.}
\end{figure}

Then, the momentum density can be divided by the normal velocity $\vv_n=-\mm^*/e^*$, which gives the normal density of the excitations. Figures \ref{fig.rhonT13} and \ref{fig.rhonfull} show the dependence of $\rho_n^{ex}$ on the normal velocity and temperature. For instance, at $T=1.3K$, the dependence of $\rho_n$ on $\vv_n$ becomes significant above $|\vv_n|>5m/s$. 

\begin{figure}[ht!]
    \centering
    \includegraphics[scale=0.5]{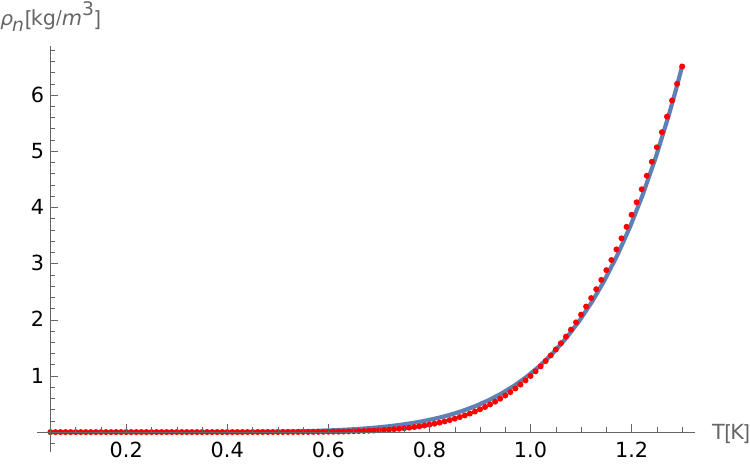}
    \includegraphics[scale=0.5]{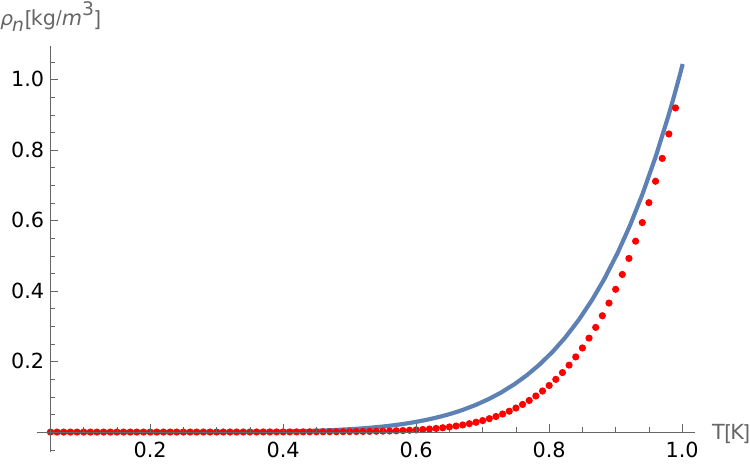}
    \caption{\label{fig.rhonfit}The dependence of the normal density $\rho_n^{ex}$ on temperature $T$ and velocity $\vv_n$. The left figure shows the dependence for the whole range of temperature, while the right figure shows the dependence for low temperatures only. Although the fit is not ideal, the normal density is approximately proportional $T^7$ and contains a term quadratic in $\vv_n$.}
\end{figure}

Figure \ref{fig.rhonfull} also shows a simple fit of the numerically obtained values,
\begin{equation}
    \rho_n^{ex} \approx \rho_{n1} T^7 + \alpha_1 T^7 |\vv_n|^2
\end{equation}
with constants $\rho_{n1} = 1.0382 kg/m^3 K^7$ and $\alpha_1 = 0.00186776 kg s^2 / m^5 K^7$. Actually, we obtain a numerical dependence of the normal density on the velocity and temperature and the above approximation is only a relatively simple fit that describes the numerical data qualitatively, see Figure \ref{fig.rhonfit}.

The analytical approximation of the normal density $\rho^{ph}_n+\rho^{rot}_n$ approximates the numerical data well in the whole range of temperatures from 0K to 1.3K. On the other hand, it would be hard to obtain the formula for the energy density from it, which is why we use the numerical fit, which also approximates the numerical data well, see Figure \ref{fig.rhonfull}.

To obtain the formula for the energy density as a function of entropy and momentum density, we start from the conjugate entropy density,
\begin{equation}
    s^* = -s + \mm^{ex}\cdot\mm^* + e\cdot e^*,
\end{equation}
which can be calculated numerically, similarly as the momentum density. The free energy density is then $F=T s^*$ and its differential reads
\begin{equation}
    \diff F = \mm^{ex}\cdot d\mm^* + e \cdot d e^* = -s dT - \mm^{ex}\cdot d\vv_n.
\end{equation}
From the approximate relation for the normal density, it follows that the free energy density has the form 
\begin{equation}
    F = F_0(T) -\frac{1}{2}\rho_{n1} T^7 |\vv_n|^2 - \frac{1}{2}\alpha_1 T^7 \vv_n^4,
\end{equation}
where $F_0(T) = -\varphi_0 T^7$ and $\varphi_0 = 350.5 J m^{-3}K^{-7}$.

\begin{figure}[ht!]
    \centering
    \includegraphics[scale=0.5]{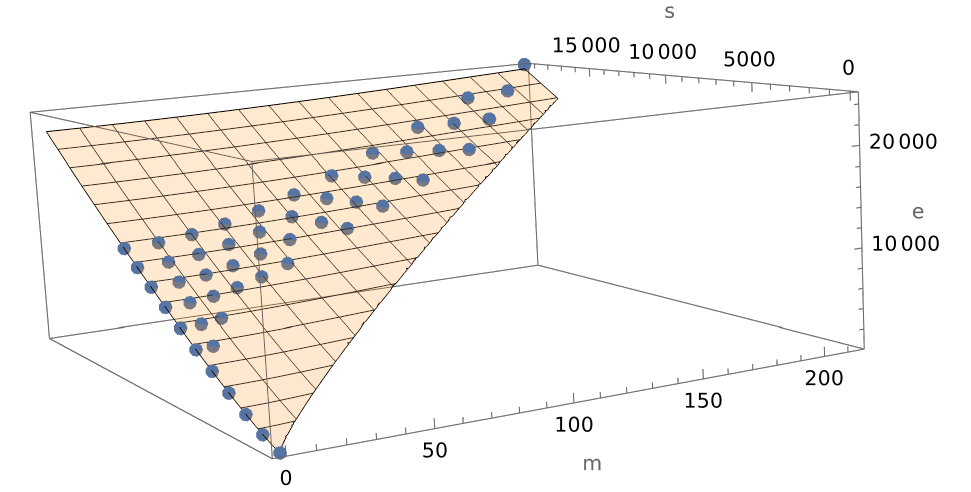}
    \caption{\label{fig.ems}The dependence of the energy density $e$ on the momentum density $\mm$ and entropy density $s$. The numerical data are obtained from the experimental dispersion relation \cite{Godfrin2021}.}
\end{figure}

The Legendre transformation to the energy density can not be carried out precisely by analytical means, but we can use the approximate relation for the normal density to obtain the energy density of excitations as 
\begin{equation}
    e\left(\mm^{ex},s\right) = e_0(\rho,s) + \frac{|\mm^{ex}|^2}{2 \rho_{n0}}+\frac{1}{4}\left(\frac{7}{24}\frac{\rho_1}{\varphi_0}-\frac{\alpha_1}{\rho_1}\right)\frac{|\mm^{ex}|^4}{\rho_{n0}^3}
\end{equation}
where $e_0=\frac{6}{7}\frac{s^{7/6}}{(7\varphi_0)^{1/6}}$, $\rho_{n0} = \rho_1 T_0^7$, and $T_0 = \left(\frac{s}{7\varphi_0}\right)^{1/6}$.
This formula approximates the numerically calculated energy density well, see Figure \ref{fig.ems}.

\subsubsection{Beyond the two-fluid model}\label{sec.beyond}
The two-fluid model of superfluid helium-4 is based on the decomposition of the total momentum density into the normal and superfluid parts, Equation \eqref{eq.m.vv}, which is based on the assumption that the energy of the excitations (Equation \eqref{eq.ep}) is quadratic in their momentum, namely the term $(\mm-\rho\vv^s)^2/2\rho_n$. 

However, in the case of both phonons and rotons, the energy is quadratic in momentum only approximately, see for instance Figure \ref{fig.rhon}. When $e^{ex}$ is not quadratic in momenta, the decomposition of the total momentum into the superfluid component and normal component, Equation \eqref{eq.m.vv}, is no longer possible. Instead, one gets a nonlinear algebraic equation by solving the phonon-entropy equation \eqref{eq.sp} for energy. In the next order of approximation, the energy \eqref{energy functional} contains, apart from the absolute and quadratic terms in the excitation momentum density, $\mm^{ex} = \mm-\rho\vv^s$, a quartic term,
\begin{align}
    e =& \frac{1}{2}\rho |\vv^s|^2 + (\mm-\rho \vv^s)\cdot\vv^s + \frac{|\mm-\rho\vv^s|^2}{2\rho_{n0}(\rho,s)}+e_0(\rho,s) - \frac{1}{4}\gamma|\mm-\rho\vv^s|^4\\
    =& \frac{1}{2}\rho |\vv^s|^2 + (\mm-\rho \vv^s)\cdot\vv^s + \frac{|\mm-\rho\vv^s|^2}{2\rho_{n}(\rho,s,\mm-\rho\vv^s)}+e_0(\rho,s) \nonumber
\end{align}
where the positive function $\gamma = \left(\frac{\alpha_1}{\rho_1}-\frac{7}{24}\frac{\rho_1}{\varphi_0}\right)\rho_{n0}^{-3}$ is identified in the above fitted model.

The normal density becomes 
\begin{equation}
    \rho_n(\rho,s,\mm-\rho \vv^s) = \frac{\rho_{n0}(\rho,s)}{1-\gamma\rho_{n0}|\mm-\rho\vv^s|^2},
\end{equation}
and it is thus not actually a function of $\rho$, $s$, and $\vv_n$, but a function of $\rho$, $s$, and $\mm-\rho\vv^s$. The normal density is thus Galilean invariant, as it should be. 

The normal velocity (the derivative of the total energy with respect to the total momentum density) then becomes
\begin{equation}
    \vv_n = \frac{\partial e}{\partial \mm} = \vv^s + \frac{\mm-(\rho_s+\rho_{n0})\vv^s}{\rho_{n0}} - \gamma |\mm-\rho\vv^s|^2(\mm-\rho\vv^s),
\end{equation}
which can be rewritten as
\begin{equation}
    \rho_{n0} \vv_n + \rho_s \vv^s = \mm - \gamma \rho_{n0} |\mm-\rho\vv^s|^2(\mm-\rho\vv^s) \neq \mm.
\end{equation}
The decomposition of the total momentum, which is essential for the two-fluid models, is valid only approximately and ceases to hold for higher velocities, higher than $5m/s$, see Figure \ref{fig.rhonT13}.

\subsubsection{Do all liquids show quantum behavior?}
As the temperature grows, the normal density approaches the value of the overall density, $\rho_n\rightarrow\rho$ \cite{sciver}, and the Hamiltonian \eqref{energy functional} gets the form of a square, 
\begin{equation}
E = \int \drr \left(\frac{|\mm|^2}{2\rho} + e_0(\rho,s)\right),
\end{equation}
which is the usual energy for classical fluids. Therefore, helium-4 behaves like a classical fluid for high temperatures. The superfluid behavior is then an effect that appears only at low temperatures, where $\rho_n < \rho$.

Imagine a classical liquid that does not freeze at temperatures sufficiently low that the energy of the motion of excitations is comparable with the internal energy. Then, would the liquid have a Hamiltonian of the form \ref{energy functional}? And would it exhibit superfluid behavior, as $\rho_n$ would no longer be equal to $\rho$?

In the above derivation, we have not used any special properties of helium-4, except for the phonon and roton dispersion relations (and the knowledge that helium-4 stays liquid). The result was that the fluid has two kinds of motion, one related to the excitations and one related to the overall (or here superfluid) motion. As the concrete form of the dispersion relation does not affect the presence of the two motions, we are tempted to ask whether any substance might exhibit superfluid behavior if it stayed liquid at low enough temperatures, regardless of the underlying microscopic properties. 

A hint towards that hypothesis lies in dilute (ultra-)cold quantum gases, where the interaction between atoms is weak. One may start with the Heisenberg equation of motion for the quantum field operator, derive a kinetic theory and even hydrodynamic equations of motion, as in the Zaremba-Nikuni-Griffin model \cite{zaremba1999}. It would be an interesting question for quantum chemistry to search for another substance that does not solidify for $T\rightarrow 0$ under ambient pressure conditions. One such example is the fermionic helium-3, whose order parameter is, however, more complicated than the scalar macroscopic wave-function of superfluid helium-4 \cite{vollhardt-wolfle}. While helium-3 is beyond the scope of this article, it should, in principle, be amenable to the approach above, albeit with a more complicated internal energy (including, e.g., order parameter textures) and the spectrum of excitations.

\subsubsection{Wave-like propagation}
The reversible part of the one-fluid model behaves like a set of wave equations, as can be seen from the Clebsch representation. In that representation, the Hamiltonian \eqref{energy functional} becomes
\begin{equation}
    E = \int \drr \left(\frac{1}{2}\rho |\nabla\rho^*|^2 + (s\nabla s^* + \alpha\nabla\alpha^*)\cdot \nabla \rho^* + \frac{\left|s\nabla s^* + \alpha\nabla\alpha^*\right|^2}{2\rho_n(\rho,s)} + e_0(\rho,s)\right).
\end{equation}
When this Hamiltonian is Taylor-expanded in the perturbations $\rho = \rho_0 + \delta\rho$, $\rho^* = \rho^*_0 + \delta\rho^*$, $s=s_0 + \delta s$, $s^* = s^*_0 + \delta s^*$, and similarly for $\alpha$, $\alpha^*$, $\beta$, and $\beta^*$, the Hamilton canonical equations for the Clebsch variables turn to a set of four wave equations, from which we can read the respective wave speeds. 

As the procedure is rather technical, let us for the sake of simplicity calculate the second-sound speed in a direct way, not based on the Clebsch representation, that is using Equations \eqref{eq.evo.MaHe.rev}. Following \cite{tilley}, we assume that the total momentum is zero and that the counter-flow velocity is small, so that the two-fluid splitting of the total momentum is valid. From the equation for the entropy density, we obtain
\begin{equation}
    \partial_t s = \nabla\cdot\left(s \frac{\rho_s}{\rho_n} \vv^s\right).
\end{equation}
From the equation for the superfluid velocity, taking into account only the term on the right-hand side that does not depend on any velocity anymore, we get that $\partial_t \vv^s \approx -\nabla \mu$, where $\mu$ is the chemical potential. Then, from the usual formula for the chemical potential, $\left(\frac{\partial \mu}{\partial T}\right)_p = -\frac{s}{\rho}$, we obtain that
\begin{equation}
    \partial_t \partial_t s = \nabla\cdot\left(\frac{s^2 \rho_s}{\rho_n \rho} \nabla \delta T\right).
\end{equation}
Further expanding the left-hand side in the small temperature perturbation $\delta T$, we get that $\partial_t s \approx \frac{1}{T_0}\rho c_p \partial_t \delta T$, where $c_p$ is the specific heat at constant pressure, $T_0$ is a constant background temperature, and $\rho$ is approximately constant. Altogether, we get the wave equation for the second sound, 
\begin{equation}
    \partial_t\partial_t \delta T = c^2_2 \nabla^2 \delta T \quad \text{with}\quad c_2^2 = \frac{s^2 T_0 \rho_s}{\rho_n \rho c_p},
\end{equation}
where $c_2$ is the second-sound speed. 
This result agrees with the literature \cite{tilley}.  

\subsubsection{Energy of quantized vortices} \label{sec: VLD and quantized vortices}
So far we have only considered the energy of the excitations and of the average motion of the atoms. In superfluid helium-4, however, there are quantized vortices that are responsible for the dissipation of the superfluid motion. The energy of quantized vortices is then added to the total energy of the fluid \cite{onsager1949,feynman1953}. 

The energy of a single vortex filament in the $z$-direction (taken as a single potential vortex with velocity $\vv^s = \frac{\hbar}{m}  \frac{1}{r} \ee_\varphi $) aligned with the unit vector in the $z$-direction, $\ee_z$, is 
\begin{equation}
    E_{\textrm{one-vortex}} = \int \drr \frac{1}{2}\rho_s |\vv^s|^2 = 
    \pi \rho_s \left(\frac{\hbar}{m}\right)^2 \ln\left(\frac{R}{a}\right) l
\end{equation}
where $R$ and $a$ are the typical vortex-boundary or vortex-vortex distance and the vortex core radius, and $l$ is the vortex length \cite{landau9}. $R$ is also often taken as the inter-vortex distance \cite{volovik-droplet}.

When there are $n$ parallel vortices in a small representative volume $V$, the overall energy then becomes
\begin{equation}\label{eq.e.vortices}
    E_{\textrm{vortices}} = \int \drr \LLL(\rr)\pi \rho_s \left(\frac{\hbar}{m}\right)^2 \ln\left(\frac{R}{a}\right)
\end{equation}
where $\LLL(\rr) = \frac{n l}{V}$ is the vortex line density in the representative volume. 
When the vortices in the representative volume are all parallel, we can replace the vortex line density $\LLL$ in the above formula with $\frac{2 \pi \hbar}{m}|\oomega|$, in accordance with \cite{Holm-HVBK}. The energy then depends on vorticity as \cite{Holm-HVBK}
\begin{equation}\label{eq.E.vortices}
	E_{\textrm{vortices}} =\int \drr  e^{\oomega}, \quad e^{\oomega} = \rho_s(\rho,s)\frac{\hbar}{2 m}\ln\left( \frac{R}{a}\right) |\oomega|,
\end{equation}
where $|\oomega|$ is the Euclidean norm of $\oomega$.
The derivative of energy with respect to the superfluid velocity becomes
\begin{subequations}
\begin{equation}
	\frac{\delta E}{\delta \vv^s} = \frac{\partial e}{\partial \vv^s} +\nabla\times \frac{\partial e^{\oomega}}{\partial \oomega}
    = \mm - \rho \vv^s -\frac{\rho}{\rho_n}(\mm-\rho\vv^s) + \nabla\times\left(\rho_s \frac{\hbar}{2m} \ln\left(\frac{R}{a}\right) \hat{\oomega}\right)
    = -\rho_s \vv_{ns} + \nabla\times\left(\rho_s \frac{\hbar}{2m} \ln\left(\frac{R}{a}\right) \hat{\oomega}\right)
\end{equation}
where $\hat{\oomega} = \oomega/|\oomega|$ is the unit vorticity vector.
The reversible part of the evolution equation for superfluid velocity \eqref{eq.vs.evorev} then gets an extra term on the right-hand side and this term determines the coefficient $B'$ in the HVBK equations, see the full equations \eqref{eq.full.explicit}.
Similarly, other derivatives of energy with respect to the state variables turn to
\begin{align}
\frac{\delta E}{\delta \rho} =& -\frac{1}{2}|\vv^s|^2 - \vv_{ns}\cdot\vv^s - \frac{1}{2}|\vv_{ns}|^2\frac{\partial \rho_n}{\partial \rho} + \mu(\rho,s) + \frac{\partial \rho_s}{\partial \rho} \frac{\hbar}{2m}\ln\left(\frac{R}{a}\right) |\oomega|\\
\frac{\delta E}{\delta s} =& -\frac{1}{2}|\vv_{ns}|^2\frac{\partial \rho_n}{\partial s} +T+ \frac{\partial \rho_s}{\partial s} \frac{\hbar}{2m}\ln\left(\frac{R}{a}\right) |\oomega|
\end{align}
while the derivative with respect to the momentum density $E_\mm = \vv_n$, given by Equation \eqref{eq.dEdm}, stays the same.
\end{subequations}

\subsection{Dissipation}\label{sec.dissipation}
So far, we have been discussing only the reversible part of the evolution equations within the one-fluid model, and now we add the irreversible (dissipative) part. 
The irreversible evolution is prescribed as gradient dynamics \cite{landau-ginzburg,dv}, being generated by a dissipation potential. Having some state variables $\xx$ and a dissipation potential $\Xi(\xx,\xx^*)$, which depends on the state variables and the conjugate variables $\xx^*$, the irreversible evolution of the state variables is given by the gradient of the dissipation potential,
\begin{equation}
    (\partial_t \xx)_{irr} = \frac{\delta \Xi}{\delta \xx^*}\Big|_{\xx^* = S_\xx}
\end{equation}
where the conjugate variables are eventually identified with the derivatives of entropy with respect to the state variables. When the dissipation potential is convex and satisfies a degeneracy condition, entropy is raised while conserving the energy \cite{kraaj,go,og}. Gradient dynamics can be seen as a framework for irreversible evolution that is consistent with the second law of thermodynamics. Another point of view is that the most probable trajectory of a stochastic system obeying a large deviation principle has the form of gradient dynamics \cite{mielke-potential}.

To better see how energy is transferred to the internal energy while raising entropy and keeping the total energy constant, we use gradient dynamics in a specific form suitable when entropy density is among the state variables \cite{dv}. This formulation gives similar results as the Single-Generator Formalism \cite{be} and can be seen as special case of generalized gradient dynamics \cite{pof2021}. For a Rayleigh dissipation potential 
\begin{equation}
    \RRR = \int \rrr(\xx, \xx^\ddagger)\dd\rr,
\end{equation}
where the energetic conjugate variables are identified with the derivatives of energy, $\xx^\ddagger = \frac{\delta E}{\delta \xx}$,
  and $\rrr$ is the density of the dissipation potential, the irreversible evolution of the state variables $\xx = (\qq,s)$ reads
\begin{subequations}\label{eq.evo.Ray}
    \begin{align}
        (\partial_t q^i)_{irr} &= -\frac{\delta \RRR}{\delta q^\ddagger_i}
        = -\frac{\partial \rrr}{\partial \qq^\ddagger_i} + \nabla\cdot\frac{\partial \rrr}{\partial \nabla q^\ddagger_i}\\
        (\partial_t s)_{irr} &= -\frac{\delta \RRR}{\delta s^\ddagger}
        + \underbrace{\frac{1}{s^\ddagger} \left(x^\ddagger_i \frac{\partial \rrr}{\partial x^\ddagger_i} + \nabla x^\ddagger_i \cdot \frac{\partial \rrr}{\partial \nabla x^\ddagger_i}\right)}_{\mbox{entropy production }\sigma}.
    \end{align}
Variables $\qq$ stand for all the state variables in $\xx$ except the entropy density.
Entropy is produced when $\RRR$ is convex and does not explicitly dependent on $s^\ddagger$ (no radiation). 
The evolution equation for energy density can be obtained by the chain rule,
    \begin{align}
        \partial_t e(\qq,s) &= -x^\ddagger_i \frac{\partial \rrr}{\partial x^\ddagger_i} + x^\ddagger_i \nabla\cdot \frac{\partial \rrr}{\partial \nabla x^\ddagger_i} + s^\ddagger \sigma\nonumber\\
        &= -\frac{\partial e}{\partial s}\frac{\partial \rrr}{\partial E_s} +  \nabla\cdot\left(x^\ddagger_i \frac{\partial \rrr}{\partial \nabla x^\ddagger_i}\right) + \nabla\cdot\left(\frac{\partial e}{\partial \nabla y^i} \nabla\cdot\frac{\partial \rrr}{\partial \nabla y^*_i} - \nabla\cdot\frac{\partial e}{\partial \nabla y^i} \frac{\partial \rrr}{\partial \nabla y^*_i}\right),
    \end{align}
    and it is conserved, unless radiation $\frac{\partial \rrr}{\partial E_s}\neq 0$ is taken into account.
    Note that $E_s$ stands for the functional derivative of energy with respect to the entropy density.
\end{subequations}

In the case of the geometric one-fluid model, the dissipation potential is assumed to have a viscous part, a heat-conductivity part, and a mutual-friction part,
\begin{subequations}
\begin{equation}
\RRR = \RRR^{(\mathrm{vis})} + \RRR^{(\mathrm{heat})} + \RRR^{(\oomega)}
\end{equation}
where
\begin{equation}
    \RRR^{(\mathrm{vis})} = \int \drr \frac{1}{2}\mu_s \nabla \mm^\ddagger :\nabla \mm^\ddagger, \qquad
    \RRR^{(\mathrm{heat})} = \int \drr \frac{1}{2}\lambda_F \nabla s^\ddagger \cdot \nabla s^\ddagger,
\end{equation}
and
\begin{equation}
    \RRR^{(\oomega)} = \int \drr \frac{\Lambda}{2}\left|\oomega\times (\vv^s)^\ddagger \right|^2 = \int \drr \frac{\Lambda}{2}\left(|\oomega|^2 |(\vv^s)^\ddagger|^2 - (\oomega\cdot(\vv^s)^\ddagger)^2\right).
\end{equation}
\end{subequations}
Coefficient $\mu_s$ is the shear viscosity and $\lambda_F$ is the Fourier heat conductivity. $\Lambda$ is a phenomenological coefficient that will be specified later in order to keep contact with the HVBK model. 
The resulting dissipative evolution of the state variables is then
\begin{subequations}\label{eq.irr}
    \begin{align}
    (\partial_t \rho)_{irr} &= 0\\
    (\partial_t \mm)_{irr} &= \nabla\cdot(\mu_s \nabla E_{\mm})\\
    (\partial_t \vv^s)_{irr} &= -\Lambda\left(\left(\oomega\times E_{\vv^s}\right)\times \oomega \right) = \Lambda\left( (\oomega\cdot E_{\vv^s})\oomega - |\oomega|^2 E_{\vv^s}\right)\\
    (\partial_t s)_{irr} &= 
        \nabla\cdot\left(\lambda_F \nabla E_s\right) + \frac{1}{E_s} \lambda_F |\nabla E_s|^2
        +\frac{1}{E_s} \Lambda\left( |\oomega|^2 |E_{\vv^s}|^2 - (\oomega\cdot E_{\vv^s})^2\right)
        +\frac{\mu_s}{E_s} \nabla E_\mm: \nabla E_\mm.
    \end{align}
\end{subequations}
Note that the total energy is conserved (the first law of thermodynamics) and that entropy is produced (the second law).

The evolution equation for superfluid velocity gets the following irreversible terms
\begin{equation}
(\partial_t \vv^s)_{irr} = -\Lambda|\oomega|^2(\vv^s)^\ddagger + \Lambda (\oomega\cdot(\vv^s)^\ddagger)\oomega =  -\frac{\rho_n}{\rho} \frac{B}{2}\hat{\oomega}\times\left(\oomega\times \left(\vv_{ns}-\nu_s \nabla\times\hat{\oomega}-\nu_s \nabla\ln(\rho_s) \times \hat{\oomega}\right)\right)
\end{equation}
for $\Lambda = \frac{B \rho_n}{\rho_s \rho |\oomega|}$, which contains the terms in Donnelly's equation for $\vv^s$ and a term dependent on $\nabla \rho_s$ \cite{donnelly1999}.

\subsection{Complete equations}
Let us now summarize the full system of equations of the geometric one-fluid model, summing the reversible (Hamiltonian) part \eqref{eq.evo.MaHe.rev} and the dissipative part \eqref{eq.irr}. The complete evolution equations for the state variables are
\begin{subequations}\label{eq.full}
\begin{eqnarray}
    \label{eq.rho.evo}\frac{\partial \rho}{\partial t} &=& -\partial_k (\rho E_{m_k}+E_{v^s_{k}})\\
    \label{eq.m.evo}\frac{\partial m_i}{\partial t} &=& 
-\partial_j(m_i E_{m_j}) -\partial_j(v^s_{i} E_{v^s_{j}}) - \rho\partial_i E_\rho -m_j \partial_i E_{m_j} - s\partial_i E_s \nonumber\\
&&-v^s_{k} \partial_i E_{v^s_{k}}+\partial_i(E_{v^s_{k}} v^s_{k})+\nabla\cdot(\mu_s \nabla E_{\mm})\\
 \label{eq.s.evo}\frac{\partial s}{\partial t} &=& -\partial_k \left(s E_{m_k}\right) +\nabla\cdot\left(\lambda_F \nabla E_s\right) + \frac{1}{E_s} \lambda_F |\nabla E_s|^2
        +\frac{\mu_s}{E_s} \nabla E_\mm: \nabla E_\mm\nonumber\\
        &&+\frac{1}{E_s} \Lambda\left( |\oomega|^2 |E_{\vv^s}|^2 - (\oomega\cdot E_{\vv^s})^2\right)\\
    \label{eq.vs.evo}\frac{\partial v^s_{k}}{\partial t} &=& -\partial_k E_\rho -\partial_k(v^s_{j}  E_{m_j}) 
+(\partial_k v^s_{j} - \partial_j v^s_{k})\left(E_{m_j}+\frac{1}{\rho}E_{v^s_{j}}\right)\nonumber\\
&& -\Lambda\left(\left(\oomega\times E_{\vv^s}\right)\times \oomega \right),
\end{eqnarray}
where energy is the sum of Galilean-transformed excitation energy \eqref{energy functional} and the energy of vortices \eqref{eq.e.vortices},
\end{subequations}
\begin{equation}\label{eq.E.tot}
E = \int \drr \left(\frac{1}{2}\rho |\vv^s|^2 + (\mm-\rho \vv^s)\cdot\vv^s + \frac{|\mm-\rho\vv^s|^2}{2\rho_n(\rho,s)}+e_0(\rho,s)
	+ \rho_s(\rho,s)\frac{\hbar}{2 m}\ln\left( \frac{R}{a}\right) |\oomega|\right).
\end{equation}
Note that we neglect the higher-order terms in the excitation momentum density, so that we stay compatible with the two-fluid model.
When we express the derivatives of energy explicitly, the evolution equations \eqref{eq.full} become a set of partial differential equations for the state variables $\rho$, $\mm$, $s$, and $\vv^s$,
\begin{subequations}\label{eq.full.explicit}
    \begin{align}
    \label{eq.rho.evo.exp}\frac{\partial \rho}{\partial t} =& -\nabla\cdot\left(\mm + \nabla\times\left(\rho_s \frac{\hbar}{2m} \ln\left(\frac{R}{a}\right) \hat{\oomega}\right)\right)\\
    \label{eq.m.evo.exp}\frac{\partial \mm}{\partial t} =& 
-\nabla\cdot(\mm \otimes \vv_{n}) -\nabla\cdot\left(\vv^s\otimes \left(-\rho_s \vv_{ns} + \nabla\times\left(\rho_s \frac{\hbar}{2m} \ln\left(\frac{R}{a}\right) \hat{\oomega}\right)\right)\right)\nonumber\\
&-\nabla \bar{P} - \nabla\cdot\left(\nabla\left(\rho_s \frac{\hbar}{2m}\hat{\oomega}\right)\times\vv^s\right)+\nabla\cdot(\mu_s \nabla \vv_n)\\
\label{eq.s.evo.exp}\frac{\partial s}{\partial t} =& -\nabla\cdot\left(s \vv_n\right) +\nabla\cdot\left(\lambda_F \nabla \bar{T}\right) + \frac{1}{\bar{T}} \lambda_F |\nabla \bar{T}|^2
        +\frac{\mu_s}{s} \nabla \vv_n: \nabla \vv_n\nonumber\\
        &+\frac{1}{\bar{T}} \Lambda\left( |\oomega|^2 \left|-\rho_s \vv_{ns} + \nabla\times\left(\rho_s \frac{\hbar}{2m} \ln\left(\frac{R}{a}\right) \hat{\oomega}\right)\right|^2 
        -\left(\oomega\cdot \left(-\rho_s \vv_{ns} + \nabla\times\left(\rho_s \frac{\hbar}{2m} \ln\left(\frac{R}{a}\right) \hat{\oomega}\right)\right)
\right)^2\right)\\
\label{eq.vs.evo.exp}\frac{\partial \vv^s}{\partial t} =& - \left(\frac{\mm}{\rho}\cdot\nabla\right)\vv^s + \frac{\mm}{\rho}\cdot \nabla \vv^s \nonumber\\
&-\nabla\left(\frac{1}{2}|\vv^s|^2 -\frac{1}{2}|\vv_{ns}|^2 \frac{\partial\rho_n}{\partial \rho} + \mu + \frac{\partial \rho_s}{\partial \rho} \frac{\hbar}{2m}\ln\left(\frac{R}{a}\right) |\oomega|\right)- \frac{1}{\rho}\oomega\times\left(\nabla\times\left(\rho_s\frac{\hbar}{2m}\ln\left(\frac{R}{a}\right)\hat{\oomega}\right)\right)\nonumber\\
&-\Lambda\left(\oomega\times \left(-\rho_s \vv_{ns} + \nabla\times\left(\rho_s \frac{\hbar}{2m} \ln\left(\frac{R}{a}\right) \hat{\oomega}\right)
\right)\right)\times \oomega \nonumber\\
=& -(\vv^s\cdot\nabla)\vv^s -\nabla\left(\mu -\frac{1}{2}|\vv_{ns}|^2 \frac{\partial\rho_n}{\partial \rho} + \frac{\partial \rho_s}{\partial \rho} \nu_s |\oomega|\right)  - \nu_s \oomega \times(\nabla\times \hat{\oomega})\nonumber\\
&-\frac{\rho_n}{\rho}\oomega \times \left(\vv_{ns} - \nu_s\nabla\times\hat{\oomega}\right) - \frac{\nu_s}{\rho}\oomega\times(\nabla\rho_s\times\hat{\oomega})\nonumber\\
&-\Lambda \rho_s \oomega \times\left(\oomega\times\left(\vv_{ns}-\nu_s\nabla\times\hat{\oomega}-\frac{\nu_s}{\rho_s}\nabla\rho_s\times\hat{\oomega}\right)\right)
\end{align}
where $\vv_n = \vv^s + (\mm-\rho \vv^s)/\rho_n(\rho,s)$, in accordance with Equation \eqref{eq.dEdm}, and $\nu_s = \frac{\hbar}{2m}\ln(R/a)$. The generalized temperature is defined as
\begin{equation}
\bar{T} = E_{s} = -\frac{1}{2}|\vv_{ns}|^2\frac{\partial \rho_n}{\partial s} +T+ \frac{\partial \rho_s}{\partial s} \frac{\hbar}{2m}\ln\left(\frac{R}{a}\right) |\oomega|.
\end{equation}
Note that the gradient in the next to last term in the equation for $\mm$ has the free index (the index equal to that of $\partial_t\mm$ on the left-hand side).
Also, for simplicity, we identify $^\flat\vv_n$ with $\vv_n$ in the above equations.
\end{subequations}

The equation for $\rho$ tells that the mass is carried by the total momentum $\mm$, as well as by the curl of superfluid vorticity. This resembles the self-induced transport in helical vortices \cite{okulov2020}. The transport of mass due to the curl of vorticity is to be subject for future research. 

The equation for the total momentum density $\mm$ is in the form of divergence of a stress tensor, so the total momentum is conserved. The stress tensor can be read as all terms under the divergence. 

The equation for entropy contains a reversible term (advection of momentum by $\vv_n$) and irreversible contributions (transport by heat flux and entropy production). 

The equation for $\vv^s$ is close to the HVBK equation for $\vv^s$ (Equation \eqref{eq.HVBK.vs}), but it contains additional terms.  From the Gibbs-Duhem relation \cite{callen}, gradient of chemical potential $\nabla\mu$ becomes identical with the $\nabla p_s$ from the HVBK equations \eqref{eq.HVBK}. On the other hand, all terms with derivatives of $\rho_n$ or $\rho_s$ in Equation \eqref{eq.vs.evo.exp} are not present in Equations \eqref{eq.HVBK}. Similar terms are present in \cite{BK1961}. Moreover, by comparison with the HVBK equations \eqref{eq.HVBK}, we obtain that $B'=2$, which is compatible with the extrapolation of experimental data \cite{barenghi1983}. 

In summary, the geometric one-fluid model is a generalization of the HVBK model, which includes the effects of vorticity and the gradient of the superfluid density. The model is consistent with the second law of thermodynamics and the conservation of energy and momentum. Moreover, it determines the value of the coefficient $B'$ in the HVBK equations.

\subsection{Limiting cases of strong and weak dissipation}
Let us now show a few limiting cases of the one-fluid model. 
Assuming that the vorticity $\oomega$ is perpendicular to the counterflow velocity, $\oomega\cdot E_{\vv^s} = 0$, Equation \eqref{eq.vs.evo.exp} simplifies to 
\begin{equation}
\frac{\partial v^s_{k}}{\partial t} = -\partial_k E_\rho -\partial_k(v^s_{j}  E_{m_j}) 
+(\partial_k v^s_{j} - \partial_j v^s_{k})\left(E_{m_j}+\frac{1}{\rho}E_{v^s_{j}}\right)
-\Lambda|\oomega|^2 E_{\vv^s}.
\end{equation}

In the limit of strong dissipation (high $\Lambda|\oomega|^2$), the equation for $\vv^s$ further simplifies to
\begin{equation}\label{eq.vs.relax}
0 = -\nabla \mu - \Lambda|\oomega|^2 E_{\vv^s},
\end{equation}
and when we plug this equation into equation for density, we obtain
\begin{equation}
\frac{\partial \rho}{\partial t} = -\nabla\cdot\left(\rho\vv_n+\frac{1}{\Lambda |\oomega|^2} \nabla\mu\right).
\end{equation}
In a superleak, where $\vv_n$ vanishes due to the viscous terms, we obtain that the mass flows into the direction of lower chemical potential $\mu$. 
As $dG = -S dT + V dp + \mu d M$ ($G$ being the total Gibbs energy in the superfluid rest frame and $M$ being the total mass), we have that $\left(\frac{\partial \mu}{\partial T}\right)_p = -\frac{s}{\rho}$ and $\left(\frac{\partial \mu}{\partial p}\right)_T = \frac{1}{\rho}$. Therefore, chemical potential decreases with temperature and increases with pressure. This explains the thermomechanical and mechanocaloric effects, where the flux of superfluid helium into hotter regions has to be compensated by increased pressure to reach the zero net flux \cite{landau6}. 

In the stationary state and when the normal velocity vanishes (like in a superleak), the $\vv^s$-equation simplifies to 
\begin{align}
\nabla \mu =& \frac{1}{\rho}E_{\vv^s}\times \oomega - \Lambda |\oomega|^2 E_{\vv^s}\\
=& \frac{\rho_s}{\rho}\vv^s \times \oomega - \rho_s\Lambda|\oomega|^2\vv^s,\nonumber
\end{align}
which is an analogue of the hydrodynamic Crocco equation \cite{Crocco}. In particular, it tells that in the case of strong dissipation, $\vv^s$ proportional to $\nabla\mu$, while in the case of small dissipation, $\vv^s$ is perpendicular to both $\oomega$ and $\nabla \mu$. 

Moreover, when no vorticity is present, the equation for $\vv^s$ simplifies to an analogue of Bernoulli equation in hydrodynamics
\begin{equation}
\frac{\vv^2_s}{2}+\mu = const,
\end{equation}
which might be used to deduce lift force acting on objects immersed in superfluid helium-4.

In the limit of strong dissipation, superfluid velocity $\vv^s$ ceases to be a dynamical variable and becomes a function of the chemical potential $\mu$. When, moreover, the timescale of observation is sufficiently long not to see any sound waves and when no significant temperature gradients are present, the liquid is also effectively incompressible and isothermal. The gradient of chemical potential then disappears and, if the vorticity is not too high, the equation $E_{\vv^s}\approx 0$ means that $\vv_n\approx\vv^s$, and the equation for $\vv^s$ may be removed from the system \eqref{eq.full}. Similarly, the equations for total density and entropy can be then removed from the system of equations, leaving only the equation for the total momentum with the generalized pressure being a scalar function determined from the condition that $\nabla\cdot\vv_n=0$. The one-fluid model then reduces to the incompressible Navier-Stokes equation. The difference between the superfluid vorticity and normal vorticity, $\nabla\times\vv_n$, then disappears, which can be used for interpretation of vorticity in experiments with optically active tracer particles \cite{Marco2021}.

\subsection{Interpretation of standard experiments}
In this section, we interpret the standard experiments in superfluid helium-4 within the one-fluid model.

\paragraph{Fountain effect \cite{tilley}:} A simplified version of the geometric one-fluid HVBK model has already been successfully used to simulate the fountain effect \cite{fountain}.
The simulation was done by modifying the Smoothed Particle Hydrodynamics (SPH) method \cite{sph}. 
Equations \eqref{eq.full} can be rewritten into a convective form with $\vv = \chi_s\vv^s + \chi_n\vv_n$, $\chi_s = \rho_s/\rho$, $\chi_n=\rho_n/\rho$
	\begin{equation}\label{eq.convective}
		\begin{split}
			\cdv{\rho} &= -\rho  \nabla \cdot \vv,\\
			\cdv{\bar{s}} &= -\frac{1}{\rho}\nabla \cdot \left(\rho \bar{s} \chi_s \vv_{ns}\right) + \frac{\beta}{\rho} \Delta T + \frac{\zeta}{\rho T}, \\
			\cdv{\vv} &= -\frac{1}{\rho}\nabla \cdot \left( \rho \chi_n \chi_s \vns \otimes \vns + p \mathbf{I} \right) + \frac{2 \mu_s}{\rho} \nabla \cdot \DD_n,\\
			\cdv{\vv^s} &=  \chi_n \nabla \vv_n ^T \vns -\frac{\nabla p}{\rho} + \bar{s} \nabla T,
		\end{split}
	\end{equation}
	where $\mu_s>0$ is the shear viscosity, $\beta >0$ is a diffusion parameter, $\bar{s}$ is the entropy per mass, and 
	\begin{equation}
		\DD_n = \frac{1}{2}(\nabla \vv_n + \nabla \vv_n^T)
  \end{equation}
  is the symmetric velocity gradient. The convective derivative is taken with respect to the overall velocity $\vv$, that is $\cdv{\bullet} = \partial_t\bullet +(\vv\cdot\nabla)\bullet$. 
  This form was then discretized into the motion of SPH particles, where each particle is equipped, apart from its position and velocity $\vv$, with the state variables $\rho$, $\bar{s}$, and $\vv^s$. 
  The geometric one-fluid model can thus recover the thermomechanical effect and gives a good agreement with the experiment.

\paragraph{Andronikashvili's experiment: }The experimental determination of the dependence of the normal fluid density on temperature carried out by Andronikashvili \cite{andronikashvili} can be also interpreted within the one-fluid model. From the convective form of the equations \eqref{eq.convective}, it follows that in the absence of the superfluid velocity, $\vv^s=0$, the momentum equation can be approximated by 
\begin{equation}
\chi_n(T) \partial_t \vv_n = \frac{2\mu_s}{\rho}\nabla\cdot \DD_n.
\end{equation}
Since the viscous stress on the right-hand side of this equation transfers the torque between the disks and helium, we can see that even in the one-fluid model the inertial forces depend on the temperature, similarly as in the two-fluid model. Instead of decomposing the liquid to the two components, the torque only initiates the normal motion, which has its own inertia, in the one-fluid model.

\paragraph{Rotating-bucket experiment:}
In the Osborne's rotating-bucket experiment \cite{osborne-bucket,tilley}, torque is transferred to Helium II via viscous forces at the boundary of the vessel. In the usual two-fluid interpretation, only the normal component rotates and forms a solid-body vortex while the superfluid component is standing still unless a critical angular velocity is reached. The critical velocity is such that the solid body rotation of the superfluid component would have the same energy as a potential vortex with the same angular momentum, and the potential (quantized) vortex is then formed \cite{landau9}. Quantized vortices then form a lattice \cite{Hall1960,tilley}. 

However, at its time, the Osborne's experiment seemed to be surprising, quoting from the abstract of \cite{osborne-bucket}, \textit{"It is shown that the results, which are the same as for any ordinary
liquid, are not in accordance with the hypothesis that the superfluid component remains
stationary"}. 

Within the one-fluid model, the experiment is interpreted as follows. The viscous forces again transfer motion to the liquid via viscous forces at the boundary, thus creating the normal motion. Due to the mutual friction in the equation for $\vv^s$, we assume that $\vv^s\approx \vv_{n}$. The equation for the total momentum density in the stationary case then simplifies to 
\begin{equation}
    0 = -\nabla p + \nabla\cdot(2\mu_s \DD_n) - \rho g \ee_z + \rho \Omega^2 r \ee_r,
\end{equation}
where gravitational and centrifugal forces were included ($g$ is the gravitational acceleration, $\Omega$ is the angular velocity of the frame, and $\ee_z$ and $\ee_r$ are unit vectors in the $z$ and $r$ directions, respectively). In the solid body rotation, when the viscous forces disappear, the pressure is given by the classical hydrostatic solution, $p = -\rho g z + \frac{1}{2}\rho\Omega^2 r^2$. The centrifugal force is then balanced by the pressure gradient. Under constant external pressure, the profile of the liquid retains the classical parabolic shape. 

In order to cover also the counter-flow experiments well, we would, however, need to include the vortex line density $\LLL$ into the state variables \cite{vinen1957,vinen1957b,vinen1957c,vinen1957d,sciacca-vinen}. A step in this direction is made in the next Section.

\section{Vortex line density} \label{sec.VLD}
Although superfluid vorticity is generally non-zero in the geometric one-fluid model, $\oomega$ does not take into account the length of vortices when they are not all parallel. Therefore, many models include also the vortex line density $\LLL$ into the state variables and thus an evolution equation is needed for it \cite{geurst1989,nemirovskii2013,Mongiovi2018}. In this Section, we show how the presence of vortex line density $\LLL(t,\rr)$ among the state variables affects the Poisson bracket of the geometric one-fluid model. 

\subsection{The Poisson bracket for the vortex line density}
The vortex line density is a volumetric density that measures the length of vortex lines in a representative volume. The Poisson bracket for the vortex line density can be then obtained by adding a scalar density to the Poisson bracket for matter in Equations \eqref{eq.PB.hydrodynamic}, which then turns to
\begin{align}\label{eq.PB.hydrodynamic.VLD}
    \{F,G\}^{(\text{hydrodynamic+L})} =& 
    \int\drr m^{ex}_i \left(\partial_j F_{m^{ex}_i}G_{m^{ex}_j} -
\partial_j G_{m^{ex}_i}F_{m^{ex}_j}\right) 
+ \int\drr s \left(\partial_i F_s
G_{m^{ex}_i} - \partial_i F_s G_{m^{ex}_i}\right)\nonumber\\
&+ \int\drr m^m_i \left(\partial_j F_{m^m_i}G_{m^m_j} -
\partial_j G_{m^m_i}F_{m^m_j}\right) 
+ \int\drr \rho \left(\partial_i F_\rho
G_{m^m_i} - \partial_i F_\rho G_{m^m_i}\right)\nonumber\\
&+ \int\drr \LLL \left(\partial_i F_\LLL G_{m^m_i} - \partial_i F_\LLL G_{m^m_i}\right).
\end{align}
When this Poisson bracket is transformed into variables $\rho$, $\mm$, $s$, $\vv^s$, and $\LLL$, it becomes
\begin{align}\label{eq.PB.HVBK.VLD}
    \{F,G\}^{(\text{HVBK}+\LLL)} &=\int\drr \rho\left(\partial_i F_\rho G_{m_i}-\partial_i G_\rho F_{m_i}\right)\\
      &\quad+ \int\drr m_i\left(\partial_j F_{m_i} G_{m_j}-\partial_j G_{m_i} F_{m_j}\right)\nonumber\\
      &\quad+ \int\drr s\left(\partial_j F_{s} G_{m_j}-\partial_j G_{s} F_{m_j}\right)\nonumber\\
    &\quad+\int\drr (G_{v^s_{i}} \partial_i F_\rho - F_{v^s_{i}} \partial_i G_\rho) \nonumber\\
&\quad+ \int\drr v^s_{j}\left(\partial_i F_{v^s_{i}} G_{m_j} - \partial_i G_{v^s_{i}} F_{m_j}\right) \nonumber\\
&\quad+ \int\drr \left(\partial_i v^s_{j} -\partial_j v^s_{i}\right) \left(F_{v^s_{i}}G_{m_j}-G_{v^s_{i}}F_{m_j}\right)\nonumber\\
&\quad+ \int \drr \frac{1}{\rho}(\partial_i v^s_{j}-\partial_j v^s_{i}) F_{v^s_{i}}G_{v^s_{j}}\nonumber\\
&\quad+ \int \drr \LLL \left(\partial_i F_\LLL G_{m_i} - \partial_i G_\LLL F_{m_i}\right)\nonumber\\
&\quad+ \int \drr \frac{\LLL}{\rho}\left(\partial_i F_L G_{v^s_{i}} - \partial_i G_L F_{v^s_{i}}\right).\nonumber
\end{align}
Poisson bracket \eqref{eq.PB.HVBK.VLD} generates reversible evolution equations
\begin{subequations}\label{eq.evo.MaHe.rev.VLD}
\begin{eqnarray}
    \label{eq.rho.evorev.VLD}\left(\frac{\partial \rho}{\partial t}\right)_{rev} &=& -\partial_k (\rho E_{m_k}+E_{v^s_{k}})\\
    \label{eq.m.evorev.VLD}\left(\frac{\partial m_i}{\partial t}\right)_{rev} &=& 
-\partial_j(m_i E_{m_j}) -\partial_j(v^s_{i} E_{v^s_{j}}) - \rho\partial_i E_\rho -m_j \partial_i E_{m_j} - s\partial_i E_s \nonumber\\
&&-v^s_{k} \partial_i E_{v^s_{k}}+\partial_i(E_{v^s_{k}} v^s_{k}) -\LLL \partial_i E_\LLL\\
 \label{eq.s.evorev.VLD}\left(\frac{\partial s}{\partial t}\right)_{rev} &=& -\partial_k \left(s E_{m_k}\right)\\
    \label{eq.vs.evorev.VLD}\left(\frac{\partial v^s_{k}}{\partial t}\right)_{rev} &=& -\partial_k E_\rho -\partial_k(v^s_{j}  E_{m_j}) 
+(\partial_k v^s_{j} - \partial_j v^s_{k})\left(E_{m_j}+\frac{1}{\rho}E_{v^s_{j}}\right) - \frac{\LLL}{\rho}\partial_i E_\LLL\\
\label{eq.L.rev} \left(\frac{\partial \LLL}{\partial t}\right) &=& -\partial_i \left( \LLL \left(E_{m_i}+\frac{1}{\rho}E_{v^s_{i}}\right)\right)
\end{eqnarray}
\end{subequations}
These equations extend the reversible evolution equations \eqref{eq.evo.MaHe.rev} by the evolution equation for the vortex line density $\LLL$. Note that subscripts stand for functional derivatives. Vortex line density is advected by the total momentum density $\mm$ in the same way as the mass density $\rho$. This is in contrast with \cite{nemirovskii2013} where $\LLL$ is advected by the counter-flow velocity $\vv_{ns}$. When $\LLL$ is advected by $\mm$, it has the advantage that the evolution equation for $\LLL$ is Galilean invariant.

Moreover, the evolution equation for $\LLL$ is coupled to both the equation for $\vv^s$ and $\mm$. Therefore, vortex line density contributes to the mutual friction force and to the stress tensor. The concrete form of those contributions is then determined by the dependence of the energy on $\LLL$. 

\subsection{Some possible dependencies of energy on vortex line density}
The dependence of energy on the vortex line density (VLD) $\LLL$ remains unclear, as a robust first-principle derivation is still missing, despite several attempts made \cite{nemirovskii2013}. For instance, if the identification of $R$ with $\LLL^{-1/2}$ were used in the expression for energy \eqref{eq.e.vortices}, the resulting energy density 
\begin{equation}
    \LLL(\rr)\pi \rho_s \left(\frac{\hbar}{m}\right)^2 \ln\left(\frac{1}{a\sqrt{\LLL}}\right)
\end{equation}
would be concave in $\LLL$ and would thus lead to unstable evolution. 

Another possibility would be to use the identification of $R$ with $\LLL^{-1/2}$ in the expression for the energy of vortices only in the logarithm. 
Following \cite{landau9}, the angular momentum of one vortex is 
\begin{equation}
  \int \drr \rho_s \vv^s\times \rr = \pi \rho_s R^2 \frac{\hbar}{m}l \ee_z
\end{equation}
and thus the angular momentum density then becomes
\begin{equation}
  \LL_{vortex} = \pi \rho_s R^2 \frac{\hbar}{m} \LLL(\rr)\ee_z.
\end{equation}
For a fluid rotating as a whole (as a solid-body vortex), the energy density is $\mathbf{\Omega}\cdot \LL$ and the angular velocity is related to the vorticity through $\oomega = 2\mathbf{\Omega}$. The energy of the rotating fluid then becomes
\begin{equation}
    \int \drr \frac{1}{2}|\oomega|\pi \rho_s R^2 \frac{\hbar}{m} \LLL(\rr),
\end{equation}
where the $\ee_z$ direction was identified with the direction of vorticity, $\oomega/|\oomega|$.
By replacing the inter-vortex distance $R$ by the vortex line density $\mathcal L$, one has
\begin{equation} \label{vortex energy both}
E_{\textrm{VLD}} 
=\int \drr  \rho_s(\rho,s)\frac{\hbar}{2 m}\ln \frac{1}{a\sqrt{\LLL(\rr)}} |\oomega|
\end{equation}
The total energy is then the sum of the energy \eqref{energy functional} and the energy of vortices, 
\begin{equation}\label{eq.E.L}
   E_{\textrm{total}} = E + E_{\textrm{VLD}}.
\end{equation}

The reversible part of the evolution equation for $L$ is obtained by the Poisson bracket \eqref{eq.PB.HVBK.VLD} and the energy functional \eqref{eq.E.L},
\begin{equation}
   \left(\frac{\partial \LLL}{\partial t}\right)_{\mathrm{rev}} = -\nabla\cdot \left( \LLL \left(\frac{\mm}{\rho}\right)\right)
    - \nabla\left(\frac{\LLL}{\rho}\right)\cdot\left(\nabla\times \left(\rho_s(\rho,s)\frac{\hbar}{2m}\ln\left(\frac{1}{\sqrt{\LLL}a}\right)\hat{\oomega}\right)\right).
\end{equation}
This equation shows, besides the advection of $\LLL$ by the average velocity $\vv$, that the evolution of vortex line density would be affected also by the superfluid vorticity $\oomega$. However, as a robust derivation of the dependence of energy on $\LLL$, based for instance on statistical mechanics, is still missing, the results of this subsection should be taken as illustrations of the possible outcomes of the Hamiltonian formulation.

The dissipative dynamics of the vortex line density is often described by the Vinen equation \cite{vinen1957c,geurst1989}. However, we do not include it here as it remains unclear how the energy depends on $\LLL$, and without that dependence it would be impossible to write equations in a thermodynamically consistent way.

A robust way towards the hydrodynamic evolution of vortex line density might be based on the nonlocal Gross-Pitaevskii equation \cite{krstulovic2023}, where vortex nucleation and roton emission can be addressed. Dissipative dynamics of the vortex line density has also been studied in the context of defocusing nonlinear Schrödinger equation \cite{berloff2007}. However, a reduction to the hydrodynamic equations, perhaps with some additional variables, is still missing.

In summary, motivated also by a critical review of the HVBK model \cite{nemirovskii2022} and the dependence of $\LLL$ on the counter-flow velocity \cite{nemirovskii-vns}, we leave the dependence of energy on $\LLL$, as well as a thermodynamic formulation of the Vinen's equation \cite{vinen1957c,Mongiovi2018} or its generalization, to future research.

\section{Conclusion}
In this manuscript, we present a geometric one-fluid model of superfluid helium-4. The model is geometrically consistent as it consists from Hamiltonian mechanics and gradient dynamics (GENERIC framework). The state variables are the total density $\rho$, total momentum density $\mm$, entropy density $s$, and superfluid velocity $\vv^s$. Neither the normal nor the superfluid density is among the state variables, so the model does not rely on the assumption that superfluid helium consist of two fluids (normal and superfluid). The Poisson bracket, generating the reversible part of the evolution, is derived from the kinetic theory of excitations and average atom motions. The energy is obtained by Galilean transformation and by statistical mechanics from an approximate and an experimental dispersion relation. The dissipative terms are then prescribed as gradient dynamics, fulfilling the first and second laws of thermodynamics while making the model compatible with the HVBK model. 

In the geometric one-fluid model, we do not assume any kind of incompressibility, which means that both normal and second sound waves can be described consistently. The superfluid velocity does not need to be curl-free and thus is to be interpreted as the average superfluid velocity within a representative volume. 

The resulting model, Equations \eqref{eq.full}, is similar to the HVBK equations. However, there are several differences. First, the mass density is advected not only by the total momentum density, but also by the curl of superfluid vorticity (terms proportional to $\nabla\times\oomega$), similar to self-induced motion of helical vortices. Second, the reversible part of the mutual friction force is determined without any phenomenological parameter (setting $B'=2$ in the HVBK equations). Eventually, we show how vortex line density may be added into the underlying Poisson bracket and how it should then be advected. 

When calculating the energy, we recover Landau's results in the lowest-order approximation in the counter-flow velocity. However, there are also higher-order terms that make the effective normal density $
\rho_n$ dependent on $\mm-\rho\vv^s$. These terms make the usual splitting of the total momentum impossible, $\mm\neq \rho_s\vv^s+\rho_n\vv_n$. The two-fluid models should be thus used carefully for higher velocities (above $5m/s$).

Finally, we show how the vortex line density can be added to the model as a state variable. The Poisson bracket for the vortex line density generates its evolution and the resulting equation is Galilean invariant. 

In the future, we would like to extend the validity of the statistical approach to temperatures above $1.3K$, taking into account also interactions between the excitations. We would also like to focus on the dependence of energy on the vortex line density, using the two-fluid kinetic theory \cite{grmela-superfluid}. Eventually, we would like to extend the derivation of the fundamental thermodynamic relation $E(\rho,s,\vv^s,\mm)$ to different pressures, so that also the dependence on density becomes explicit.

\section*{Acknowledgment}
NC and MP were supported by Czech Science Foundation, project 23-05736S. MP is a member of the Nečas centre for mathematical modeling. EV was supported by the Charles University under PRIMUS/23/SCI/017. We are grateful to D. Schmoranzer, M. La Mantia, and L. Skrbek for invaluable discussions on superfliud helium-4. Also, we would like to thank M. Grmela, D. Jou, and L. Restuccia for dicussing the geometry and non-equilbirium thermodynamics of superfluids. 

\appendix
\section{Quantum--classical correspondence} \label{subsec:correspondence}
Although superfluid helium-4 is a system with quantum properties (for instance the presence of quantized vortices and persistent currents \cite{tilley}), there does not exist a complete quantum theory for describing its dynamics. As emphasized in Section \ref{sec:model comparison}, most models are interpreted as classical hydrodynamic models. However, as there is a hierarchy between Poisson brackets, one can, for the construction of a model, always start from the canonical Poisson bracket between position $r^j$ and momentum $p_j$,
\begin{equation} \label{eq: PB r,p}
\{r^i, p_j\} = \delta^i_j.
\end{equation}

The commutator between the position operator $\hat r_b$ and momentum operator $\hat r_a$ is known to be given by 
\begin{equation} \label{eq:comm r,p}
[\hat p_a , \hat r_b] =i\hbar \delta_{ab},
\end{equation}
and the canonical Poisson bracket \ref{eq: PB r,p}, is related to the commutator \eqref{eq:comm r,p} by 
\begin{equation} \label{relation comm PB}
\frac{1}{i\hbar }[\hat p_a , \hat r_b] =\{r^b, p_a\},
\end{equation}
see for instance \cite{landau3,ma67}.

The temporal evolution of observables, be they hydrodynamic quantum operators or classical hydrodynamic variables, can be derived in the Heisenberg representation (\cite{Landau1941, dv}), or, respectively, by applying the Correspondence Principle \eqref{relation comm PB}, from the Poisson bracket. In any way, this is possible if the underlying theory is Hamiltonian.

On the other hand, Groenewald showed that the Correspondence Principle does not work for nonquadratic Hamiltonians \cite{groenewold46}. The evolution equations for the average observed values do not coincide with the corresponding classical evolution equations. Therefore, our goal is not to obtain the full set of evolution equations by the Correspondence Principle, but we only focus on the Poisson bracket. Actually, the Poisson bracket is completely specified by its action on linear functionals \cite{fecko}, where the Groenewald argument does not hold. 

For the formulation of the geometric one-fluid model, one can start from the canonical commutation relation \eqref{eq:comm r,p}, apply the Correspondence Principle, and derive the noncanonical Poisson bracket and in Section \ref{sec.PB}. Alternatively, the bracket may be obtained as the Group Poisson bracket (PB) method \cite{dv}. This method uses the relation between the commutator of a Lie algebra and the functional derivative
\begin{equation} \label{VoloviKPB}
\lambda [L^k(x),A(y)]=\frac{\delta A(y)}{\delta \alpha ^k(x)},
\end{equation}
where $L^k(x)$ is a generator of a unitary transformation. For instance, the momentum operator $\hat p_k$, which generates translations $u_k$ is space according to $x_k\rightarrow x_k+u_k$, $A(y)$, is a functional depending on spacial variables $y$, and $\alpha_k(x)$ is the effect of the unitary transformation, e.g. $u_k$. The method assumes that a formal relation \eqref{VoloviKPB} is also valid for any Poisson bracket (as the algebra of the quantum commutators and Poisson brackets of classical analytical mechanics share basically all formal properties). Hence, Poisson brackets can be derived by analogy to the quantum theory \eqref{VoloviKPB} according to 
\begin{equation} \label{VoloviKPB classical}
\{l^k(x),l^m(y)\}=\frac{\delta l^k(y)}{\delta \alpha ^k(x)}.
\end{equation}
The hydrodynamic Euler equations can be derived this way, as well as the Landau-Tisza hydrodynamic two--fluid model of helium-4 \cite{dv}. Note that it is not necessary to know the explicit form of the Hamiltonian when deriving only the Poisson bracket.

\end{document}